\documentclass[nofootinbib,prd]{revtex4}

\usepackage{times,epsfig}
\usepackage{graphics,graphicx}
\usepackage{amsmath}
\usepackage{amsfonts}
\usepackage{epsfig}
\usepackage{color}
\input{colordvi.tex}
\usepackage{braket}
\usepackage{yfonts}

\usepackage{enumerate}

\usepackage{mathrsfs}
\usepackage{amssymb}
\usepackage{bm}

\def\lsim{\,\lower2truept\hbox{${<\atop\hbox{\raise4truept\hbox{$\sim$}}}$}\,}
\def\gsim{\,\lower2truept\hbox{${> \atop\hbox{\raise4truept\hbox{$\sim$}}}$}\,}
\def\simlt{\mathrel{\rlap{\lower 3pt\hbox{$\sim$}}
        \raise 2.0pt\hbox{$<$}}}
\def\simgt{\mathrel{\rlap{\lower 3pt\hbox{$\sim$}}
        \raise 2.0pt\hbox{$>$}}}

\def\be{\begin{equation}}
\def\ee{\end{equation}}
\def\ba{\begin{eqnarray}}
\def\ea{\end{eqnarray}}

\def\DHLhksqrt#1#2{\setbox0=\hbox{$#1\sqrt{#2\,}$}\dimen0=\ht0
\advance\dimen0-0.2\ht0
\setbox2=\hbox{\vrule height\ht0 depth -\dimen0}%
{\box0\lower0.4pt\box2}}

\definecolor{darkred}{RGB}{175,0,0}
\definecolor{darkblue}{RGB}{14,0,185}

\begin{document}

\title{Future Constraints on Angle-Dependent Non-Gaussianity from Large Radio Surveys}

\author{Alvise Raccanelli$^{1}$,  Maresuke Shiraishi$^{2,3,4}$, Nicola Bartolo$^{2,3}$, Daniele Bertacca$^{5}$, Michele Liguori$^{2,3}$, Sabino Matarrese$^{2,3,6}$, Ray P. Norris$^{7}$, David Parkinson$^{8}$ \\~}

\affiliation{
$^{1}$ Department of Physics \& Astronomy, Johns Hopkins University, 3400 N. Charles St., Baltimore, MD 21218, USA \\
$^{2}$ Dipartimento di Fisica e Astronomia ``G. Galilei'', Universit\'a degli Studi di Padova, via Marzolo 8, I-35131, Padova, Italy \\
$^{3}$ INFN, Sezione di Padova, via Marzolo 8, I-35131, Padova, Italy \\
$^{4}$ Kavli Institute for the Physics and Mathematics of the Universe (Kavli IPMU, WPI), UTIAS, The University of Tokyo, Chiba, 277-8583, Japan \\
$^{5}$ Argelander-Institut f\"ur Astronomie, Auf dem H\"ugel 71, D-53121 Bonn, Germany \\
$^{6}$ Gran Sasso Science Institute, INFN, viale F. Crispi 7, 67100 L'Aquila, Italy \\
$^{7}$ CSIRO Astronomy and Space Science, Australia Telescope National Facility, PO Box 76, Epping NSW 1710, Australia \\
$^{8}$ School of Mathematics and Physics, University of Queensland, Brisbane, QLD 4072, Australia
}
\date{}

\begin{abstract}
We investigate how well future large-scale radio surveys could measure different shapes of primordial non-Gaussianity; in particular we focus on angle-dependent non-Gaussianity arising from primordial anisotropic sources, whose bispectrum has an angle dependence between the three wavevectors that is characterized by Legendre polynomials $\mathcal{P}_L$ and expansion coefficients $c_L$.
We provide forecasts for measurements of galaxy power spectrum, finding that Large-Scale Structure (LSS) data could allow measurements of primordial non-Gaussianity competitive or improving upon current constraints set by CMB experiments, for all the shapes considered.
We argue that the best constraints will come from the possibility to assign redshift information to radio galaxy surveys, and investigate a few possible scenarios for the EMU and SKA surveys. A realistic (futuristic) modeling could provide constraints of $f_{\rm NL}^{\rm loc} \approx 1 (0.5)$ for the local shape, $f_{\rm NL}$ of $\mathcal{O}(10) (\mathcal{O}(1))$ for the orthogonal, equilateral and folded shapes, and $c_{L=1} \approx 80 (2)$,  $c_{L=2} \approx 400 (10)$ for  angle-dependent non-Gaussianity.
The more futuristic forecasts show the potential of LSS analyses to considerably improve current constraints on non-Gaussianity, and so on models of the primordial Universe.
Finally, we find the minimum requirements that would be needed to reach $\sigma(c_{L=1})=10$, which can be considered as a typical (lower) value predicted by some (inflationary) models.
\end{abstract}


\date{\today}

\maketitle


\section{Introduction}
\label{sec:intro}
Investigating models describing the primordial universe is a priority for current and planned cosmological experiments; recent results from measurements of the CMB by e.g. the {\it  Planck} satellite~\cite{Adam:2015rua,Ade:2013uln,Ade:2015lrj,plancknG, plancknG2015} and of the Large-Scale Structure (LSS) of the Universe, by e.g. the SDSS~\cite{Ross:2013, Agarwal:2014, Leistedt:2014} already set some constraints on such models, but more precision is needed in order to further discriminate between them.
The window for investigating the inflationary period is given by the statistics of primordial curvature fluctuations, and in particular the (deviations from) gaussianity of the probability distribution function of primordial cosmological perturbations. Imprints of primordial non-Gaussianity in the CMB anisotropies and in the LSS are usually described in terms of shapes (e.g.~\cite{Babich:2004gb, Bartolo:2004if, Liguori:2010hx, Chen:2010xka, Komatsu:2010hc, Yadav:2010fz}).
Recently, along with ``standard'' shapes like the local, equilateral, orthogonal and folded templates, shapes with specific angle dependence have also been investigated (e.g., \cite{Barnaby:2012tk, Shiraishi:2012rm, Shiraishi:2012sn, Bartolo:2012sd, Endlich:2012pz, Shiraishi:2013vja, Endlich:2013jia}).

The angle dependence of primordial non-Gaussianity can be generated by the existence of non-Gaussian anisotropic sources. We may parameterize the angular dependence in the bispectrum of curvature perturbations using Legendre polynomials $\mathcal{P}_L(x)$~\cite{Shiraishi:2013vja}:
\begin{eqnarray}
\label{BL}
B_\zeta^{L}(k_1, k_2, k_3) = c_L \left[ \mathcal{P}_L(\hat{\bf k}_1 \cdot \hat{\bf k}_2) P_\zeta(k_1) P_\zeta(k_2) + 2~{\rm perm} \right] \label{eq:zeta3_cL}~,
\end{eqnarray}
where $c_0$ is related to the local-type non-Gaussianity parameter as $c_0 = (6/5) f_{\rm NL}^{\rm loc}$. Non-Gaussian anisotropic sources actually create nonzero $c_L$ for $L \geq 1$. For example, in a model where the inflaton field $\phi$ is coupled to a U(1) gauge field via a $f(\phi) F^2$ interaction (where $F_{\mu \nu}=\partial_\mu A_\nu-\partial_\nu A_\mu$ is the vector field strength), non-vanishing $c_2$ arise as well as $c_0$, whose magnitudes are related to each other as $c_2 = c_0 /2$~\cite{Shiraishi:2013vja,Bartolo:2012sd}.~\footnote{Notice in the $f(\phi) F^2$ where the vector field has a non-vanishing vev a statistical anisotropic bispectrum is actually generated and, after an angle-average, it takes the form~(\ref{BL}). For studies of bispectra that break statistical isotropy see~\cite{Bartolo:2011ee,Shiraishi:2011ph}.} The case of $c_2 \gg c_0$ can be realized in the so-called solid inflation models~\cite{Endlich:2012pz, Bartolo:2013msa, Endlich:2013jia, Bartolo:2014xfa}, which are based on a specific internal symmetry obeyed by the inflaton fields.  In the presence of large-scale non-helical and helical magnetic fields in the radiation-dominated era, not only $c_0$ and $c_2$ but also $c_1$ can be created~\cite{Shiraishi:2012rm, Shiraishi:2012sn, Shiraishi:2013vja}.
Recently a model with a $f(\phi) (F^2 + F\tilde{F})$ coupling has been proposed as the first clear example of an inflationary model where $c_1$ is generated,  realizing $c_0 : c_1 : c_2 = 2 : -3 : 1$ \cite{Bartolo:2015dga}.

Therefore, $c_1$ and $c_2$ can become key observables to probe these models. So far, observational constraints have been investigated via the CMB bispectrum analysis. In an all-sky ideal measurement of the temperature anisotropies up to $\ell = 2000$, where instrumental noise is completely negligible, expected $1\sigma$ error bars on $c_1$ and $c_2$ reach $61$ and $13$, respectively~\cite{Shiraishi:2013vja}. Realistic constraints have been obtained from the {\it Planck} temperature data, giving $c_1 = 118 \pm 103$ and $c_2 = - 5 \pm 26$ (68\% CL).
\footnote{These values correspond to the {\it Planck} 2015 limits~\cite{plancknG2015}: $f_{\rm NL}^{L=1} = -49 \pm 43$ and $f_{\rm NL}^{L=2} = 0.5 \pm 2.7$, where the $f_{\rm NL}^L$ parameters in the {\it Planck} papers \cite{plancknG, plancknG2015} are related to $c_L$ in Eq.~(\ref{eq:zeta3_cL}) as $c_1 = - (12/5) f_{\rm NL}^{L=1}$ and $c_2 = - (48/5) f_{\rm NL}^{L=2}$.}

As well as in the CMB~\cite{Bartolo:2004if, Komatsu:2010, plancknG, plancknG2015}, signatures of primordial non-Gaussianity can be also measured in the LSS of the Universe via the scale dependence of the halo bias (see e.g.~\cite{Matarrese:2000, Matarrese:2008, Dalal:2008}). In this paper, we examine how powerful can LSS surveys be in measuring the angle-dependent non-Gaussianity parameters $c_1$ and $c_2$ which could reveal about specific details of the physics of inflation, such as the presence of vector fields.
\footnote{This paper focuses on the isotropic angle-averaged bispectrum effects on the scale-dependent bias due to primordial vector fields. See \cite{Baghram:2013lxa} for the analysis including the statistically-anisotropic effects. For the analysis of the effects sourced from large-scale vector fields existing in the radiation dominated era, see~\cite{Shiraishi:2013sv}.}

Galaxy clustering can and has been used to set constraints on e.g. dark energy parameters~\cite{Samushia:2011}, models of gravity~\cite{Raccanelligrowth}, neutrino mass~\cite{dePutter:2012, Zhao:2012}, dark matter models~\cite{Cyr-Racine:2014, Dvorkin:2014}, the growth of structures~\cite{Samushia:2013, Reid:2012}, and here we are in particular interested in their use to measure primordial non-Gaussianity.
Non-Gaussianity parameters can be measured via the LSS by using the Integrated Sachs-Wolfe effect~\cite{Xia:2010, Raccanelli:2015ISW} and the galaxy power spectrum (see e.g.~\cite{Matarrese:2000, Dalal:2008, Slosar:2008, Desjacques:2011, Xia:2010, Raccanelli:2014fNL, Ho:2013, SKA:Camera, SKA:Abdalla, SKA:Jarvis}; for some forecasts on measurements on non-Gaussianity with 21-cm surveys, see e.g.~\cite{Tashiro:2012, DAloisio:2013, Mao:2013, Camera:2013, Munoz:2015}. For a recent review on testing inflation with galaxy surveys, see in particular~\cite{dePutter:2014, Alvarez:2014}).

Recent measurements from LSS include~\cite{Xia:2010, Ross:2013, Agarwal:2014, Leistedt:2014}; with the current generation of experiments, these constraints are not competitive with CMB ones, but future galaxy surveys will provide extremely precise measurements of galaxy clustering over a wide area of the sky and a large range in redshifts (see~\cite{Font-Ribera:2013, PFS, SKA}.
For this reason, planned experiments such as SPHEREx~\cite{spherex} will be explicitly focused on investigating the primordial universe (see also~\cite{Raccanelli:2014fNL, dePutter:2014} for a general overview of expected constraints).

The paper is organized as follows.
In Section~\ref{sec:vector} we study the effects of angle-dependent non-Gaussianity on the halo bias and compare them to effects due to other shapes. In Section~\ref{sec:surveys} we present the survey specifications and assumptions and in Section~\ref{sec:pk} we describe the observable used to compute our forecasts.
In Section~\ref{sec:results} we present the methodology adopted for obtaining our results and we present  them, with a focus on future constraints on vector fields models in Section~\ref{sec:vec_con}.
Finally, in Section~\ref{sec:conclusions} we draw our conclusions.


\section{Halo bias in the presence of angle-dependent non-Gaussianity}
\label{sec:vector}
It is well-known that non-vanishing non-Gaussianity can create nontrivial scale dependence in the bias parameter associating halos with matter fluctuations, in addition to a usual scale-invariant contribution in the Gaussian case $b_{\rm G}$. In the presence of nonzero primordial non-Gaussianity, one may write the total bias as:
\begin{equation}
\label{eq:ng-bias}
b_{\rm NG}(z, k) = b_{\rm G}(z) + \Delta b(z, k) ~.
\end{equation}

There are diverse approaches to estimating the scale-dependent part $\Delta b$ for given primordial non-Gaussianities (see e.g.~\cite{Matarrese:2008, Dalal:2008, Slosar:2008, Desjacques:2011, Xia:2010, Matsubara:2012nc}). In this work, we estimate $\Delta b$ by means of the iPT formalism \cite{Matsubara:2012nc}, which is applicable to the analysis of almost any type of primordial non-Gaussianity. The corresponding formula reads:
\begin{eqnarray}
\Delta b(z, k) \approx \frac{\sigma_M^2}{2 \delta_c^2} 
\left[ A_2(M) {\cal I}(k, M) + A_1(M) \frac{\partial {\cal I}(k, M)}{\partial \ln \sigma_M} \right], \label{eq:bias}
\end{eqnarray}
where $\delta_c = 1.686$ is the critical overdensity and $\sigma_M$ denotes the density variance per the mass of halo, $M \equiv \frac{4\pi}{3} \rho_{m0} R^3$:
\begin{eqnarray}
\sigma_M^2 = \int \frac{k^2 d k}{2 \pi^2} W^2(kR) P_{\rm L}(k) \, ,
\end{eqnarray}
$\rho_{m0}$ being the mean matter density at the present epoch and $P_{\rm L}(k)$ the linear matter power spectrum. \footnote{In this paper we only take into account the contributions from the bispectrum. See \cite{Baumann:2012bc, Yokoyama:2012az} for discussions on the higher-order effects.} As a window function, we adopt a top-hat form: $W(x) = \frac{3 j_1(x)}{x}$. The coefficients $A_i(M)$ are determined by the mass function of halos. In the numerical works of this paper, we use fitting formulae for $A_i(M)$ based on the MICE mass function \cite{Crocce:2009mg}, whose explicit forms are derived in~\cite{Matsubara:2012nc}.

The function ${\cal I}$ is given by a convolution of the linear matter bispectrum $B_{\rm L}$ (filtered with the window function):
\begin{eqnarray}
\label{eq:I_function}
{\cal I}(k, M) = 
\frac{1}{\sigma_M^2 P_{\rm L}(k)}  
\int \frac{d^3 {\bf k'}}{(2 \pi)^3} W(k' R) W(|{\bf k} - {\bf k'}| R) 
 B_{\rm L}(k, k', |{\bf k} - {\bf k'}|) 
~.
\end{eqnarray}
In linear perturbation theory, matter fluctuations become a simple product of primordial curvature perturbation and the linear matter transfer function, i.e. $\delta({\bf k}, z) = {\cal M}_\zeta(k,z) \zeta_{\bf k}$, and hence their power spectrum and bispectrum at given $z$ are written respectively as:
\begin{equation}
P_{\rm L}(k) = {\cal M}_\zeta^2(k,z) P_{\zeta}(k) \, , \,\,\,\,\,\,\,
B_{\rm L}(k_1, k_2, k_3) = \left[\prod_{n=1}^3 {\cal M}_\zeta(k_n,z) \right] B_{\zeta}(k_1, k_2, k_3) \, .
\end{equation}
In the standard $\Lambda$CDM cosmology: 
\begin{equation}
{\cal M}_\zeta(k,z) = \frac{2}{5}D(z)\frac{k^2 T(k)}{H_0^2 \Omega_{m0}} \; ;
\end{equation}
a derivation can be found in~\cite{Shiraishi:2013sv, Lahav:1991, Weinberg}. From these relations, one can deduce that the $k$ dependence of $\Delta b$ varies according to the shape of the primordial bispectrum $B_\zeta$, via the variation of the function ${\cal I}$. In the next section, by analyzing matter power spectra and the bias parameter, we will explore the detectability of angle-dependent non-Gaussianity, i.e., $c_1$ and $c_2$ of Eq.~(\ref{eq:zeta3_cL}), as well as the detectability of the usual four types of non-Gaussianity, i.e., $f_{\rm NL}^{\rm loc}$, $f_{\rm NL}^{\rm eq}$, $f_{\rm NL}^{\rm ort}$ and $f_{\rm NL}^{\rm fol}$. The data of $\Delta b$ used in such analysis are obtained through numerical computations of the iPT formalism as in Eq.~(\ref{eq:bias}). 

\begin{figure*}[t]
\includegraphics[width=0.49\linewidth]{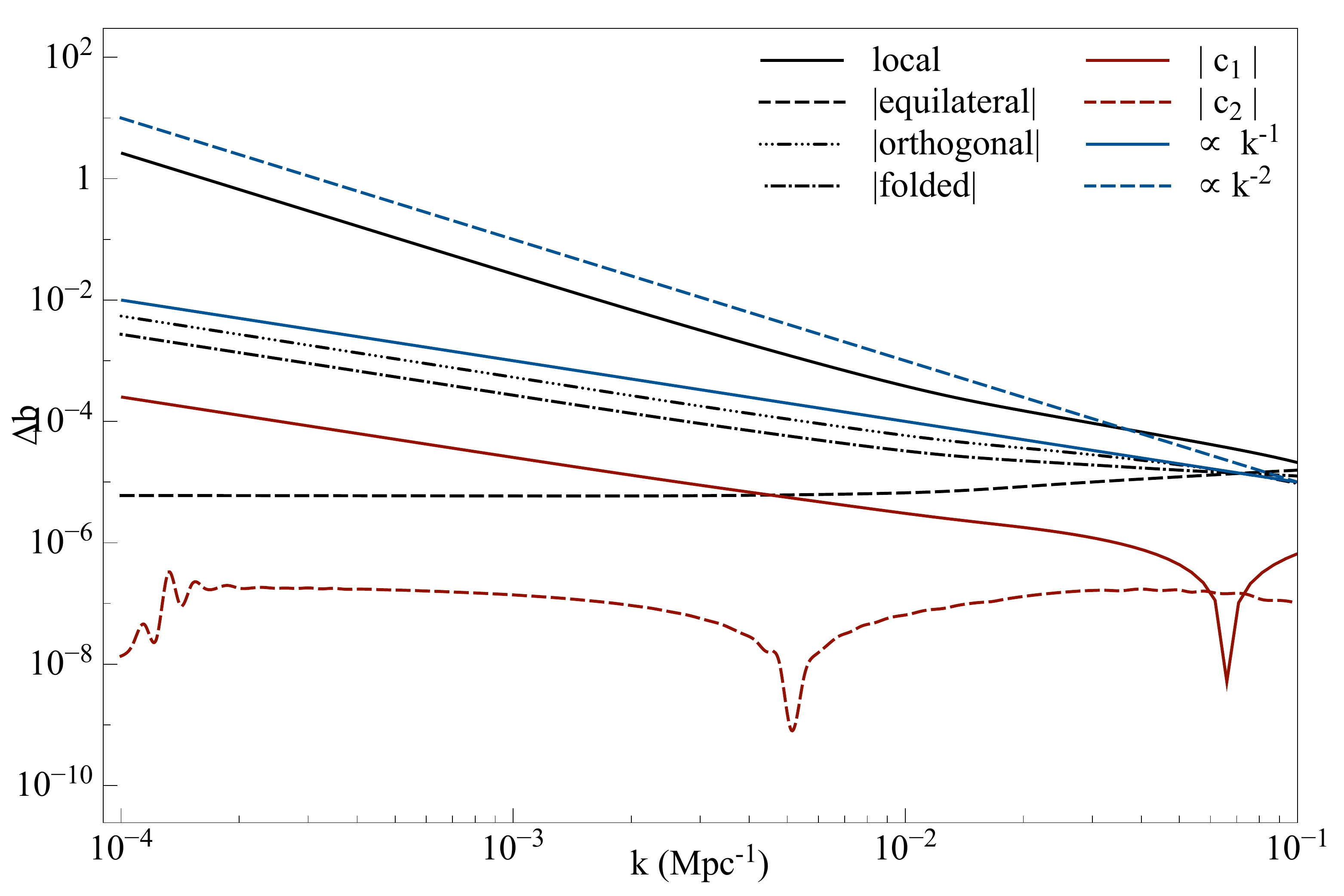}
\includegraphics[width=0.49\linewidth]{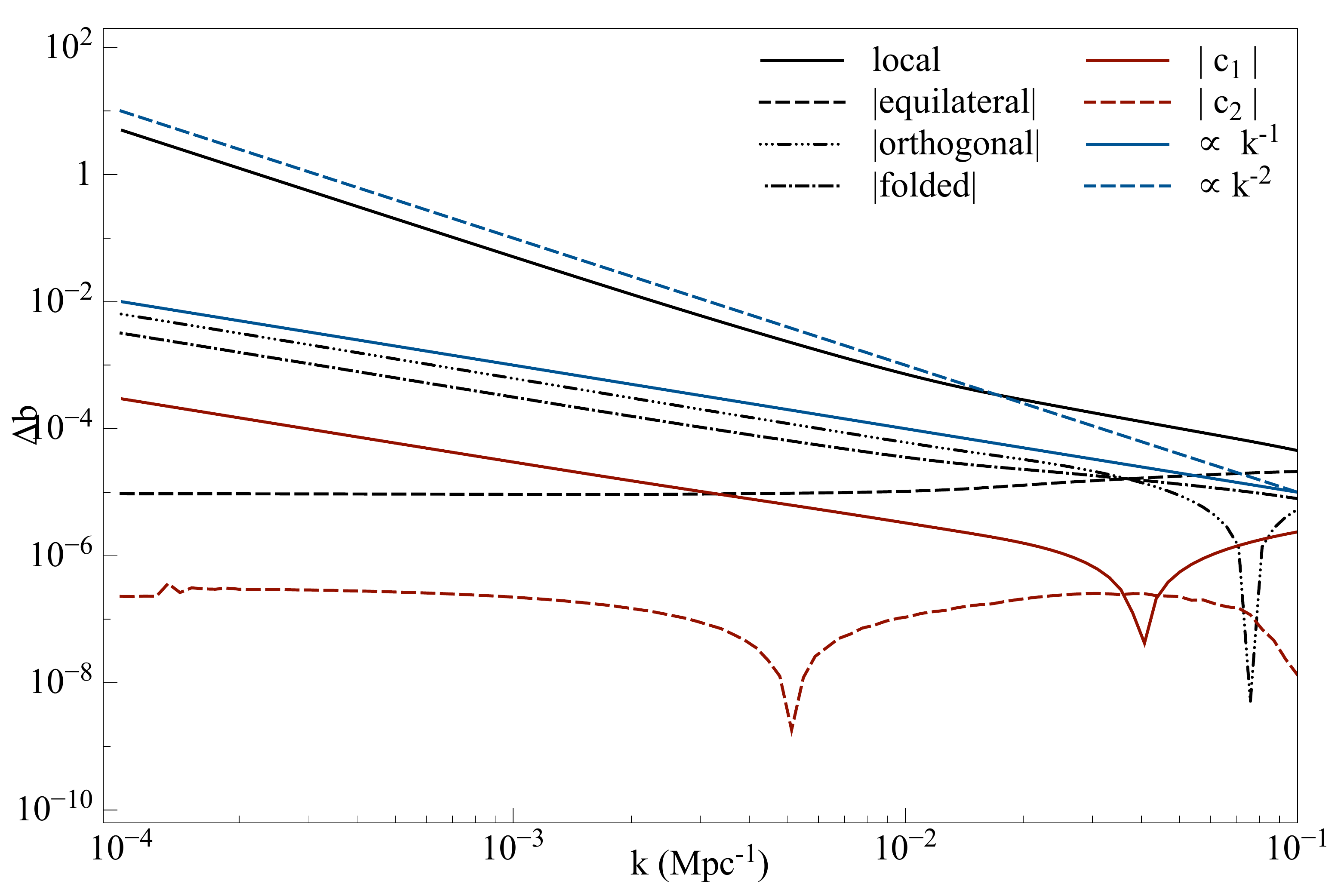}
\includegraphics[width=0.49\linewidth]{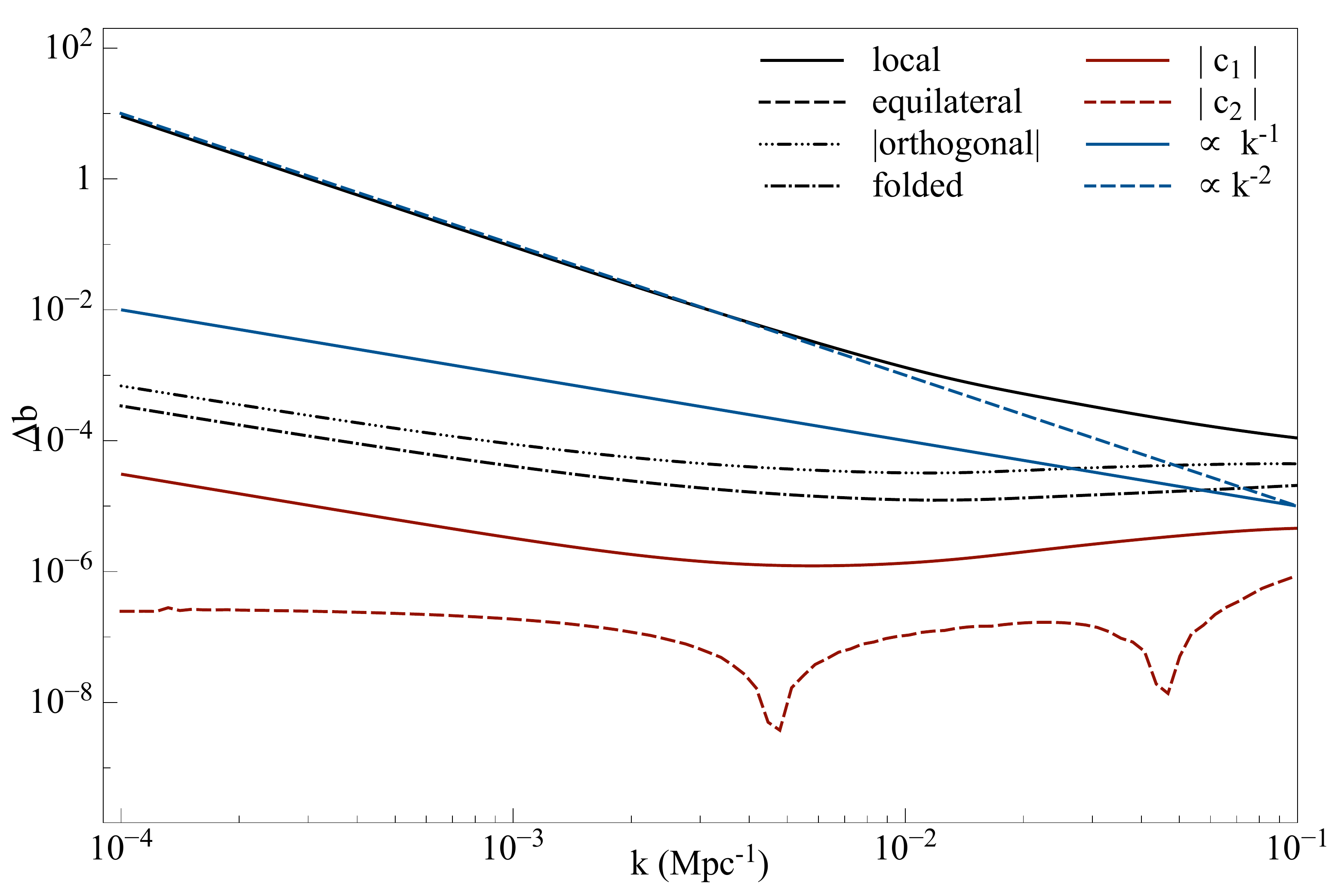}
\includegraphics[width=0.49\linewidth]{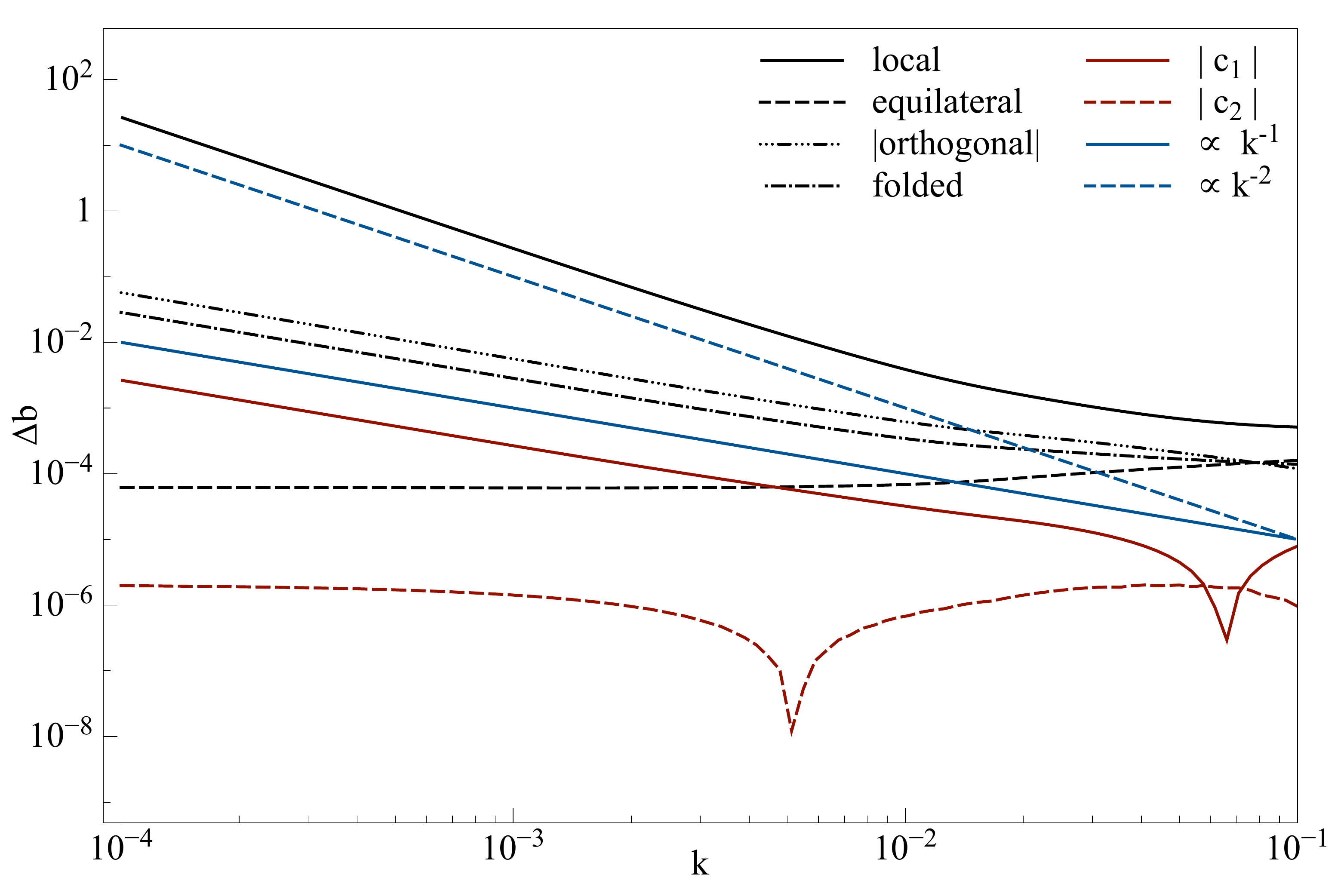}
\caption{Scale-dependent correction to the bias, for different shapes of non-Gaussianity at $z=1$. Here we fix the non-linearity parameters as $f_{\rm NL}^{\rm loc} = f_{\rm NL}^{\rm eq} = f_{\rm NL}^{\rm ort} = f_{\rm NL}^{\rm fol} = c_1 = c_2 = 1$, and adopt $M = 10^{11} M_\odot$ (top left panel), $M = 10^{12} M_\odot$ (top right panel), $M = 10^{13} M_\odot$ (bottom left panel) and $M = 10^{14} M_\odot$ (bottom right panel).
}
\label{fig:bias_z1}
\end{figure*}

Our numerical results for $\Delta b(z,k)$ for each primordial non-Gaussianity model at $z = 1$ are shown in Figure~\ref{fig:bias_z1}. It is visually apparent there that the tilt of $k$ varies based on the non-Gaussianity shape. Some simple scaling arguments can help in understanding the main features of the results of Figure~\ref{fig:bias_z1}. It is a known fact that the convolution in the ${\cal I}$ function is determined, in the large-scale limit, by the behavior of $B_{\rm L}$ in the squeezed configurations, $k \ll k'\simeq |{\bf k} - {\bf k}'|$. In particular for the squeezed-limit of the the angle-dependent primordial bispectrum we find:
\begin{eqnarray}
B_\zeta^{L=0}(k,k',|{\bf k} - {\bf k}'|) & \to & 2 c_0 P_\zeta(k) P_\zeta(k') \propto B_\zeta^{\rm loc}(k,k',k') \,, \nonumber \\ 
B_\zeta^{L=1}(k,k',|{\bf k} - {\bf k}'|) & \to & - c_1 \frac{k}{k'} P_\zeta(k) P_\zeta(k') \nonumber \\
B_\zeta^{L=2}(k,k',|{\bf k} - {\bf k}'|) & \to & 2 c_2 {\cal P}_2(\hat{\bf k} \cdot \hat{\bf k}') P_\zeta(k) P_\zeta(k') 
\,.
\label{eq:zeta3_c0_c1_squeezed}
\end{eqnarray}
Note that for the $L=1$ case, taking the squeezed limit, the explicit angle dependence ${\cal P}_1(\hat{\bf k} \cdot \hat{\bf k}')$ vanishes because of the property of odd function, and only a ${\cal O}(\frac{k}{k'})$ term remains. Taking into account the squeezed limit of the other bispectra considered in this paper,  the large-scale behavior of the transfer function, ${\cal M}_\zeta \propto k^2$, and assuming scale invariance of $P_\zeta$ and $B_\zeta$, we analytically estimate the following scaling relations on large scales as: \footnote{
Notice that for the equilateral, orthogonal and folded shapes we are using the standard template shapes~\cite{Babich:2004gb,Senatore:2009gt}. In fact these templates do not recover the precise squeezed limit of the theoretical models on which they are based, a limit to which the scale dependent bias can be particularly sensitive to (see, e.g. discussions in~\cite{Creminelli:2010qf,Norena:2012yi} Despite of this we have used these templates, both because our main goal is to focus on the angle-dependent non-Gaussianity, and also to easily compare with previous results in the literature (the use of the standard templates can be anyway taken as examples of the efficiency to use the scale dependent halo bias to constrain these shapes). On the contrary, for the scale dependent non-Gaussianity we are using the bispectra predicted by the underlying physical model(s), including the large-scale scaling $k^{-1}$ for the $c_1$ shape in the squeezed limit.}

\begin{eqnarray}
\Delta b \propto
\begin{cases}
k^{-2} &: {\rm local \ or \ }  c_0  \\
k^0 &: {\rm equilateral} \\
k^{-1} &: {\rm orthogonal} \\
k^{-1} &: {\rm folded} \\
k^{-1} &: c_1 
\end{cases} \label{eq:deltab_scale}
\end{eqnarray}
Figure~\ref{fig:bias_z1} shows that our numerical results match these analytic expectations (on large scales). Such estimation also yields $\Delta b \to 0$ for the $c_2$ case, because an angular integral in the convolution~\eqref{eq:I_function} is reduced to $\int_{-1}^1 d\mu {\cal P}_2(\mu)$ in the squeezed limit and this becomes zero. This essentially explains the smallness of our $c_2$ result described in Figure~\ref{fig:bias_z1}.
The large-scale amplitude of $\Delta b$ in the angle-dependent $c_{L = 0, 1, 2}$ models decreases as the Legendre multipole $L$ increases, while, interestingly, they are close to each other on very small scales, due to the difference of their scaling as $\Delta b^{L=0} \propto k^{-2}$, $\Delta b^{L = 1} \propto k^{-1}$ and $\Delta b^{L = 2} \propto k^0$. 
We will investigate the observational consequences in the rest of the paper, and in particular in Section~\ref{sec:vec_con}.
Our results for the standard four shapes are also in excellent agreement with previous studies~\cite{Matsubara:2012nc, Byun:2014}, confirming the validity of our numerical computations. 

\begin{figure*}[thb!]
\includegraphics[width=0.49\linewidth]{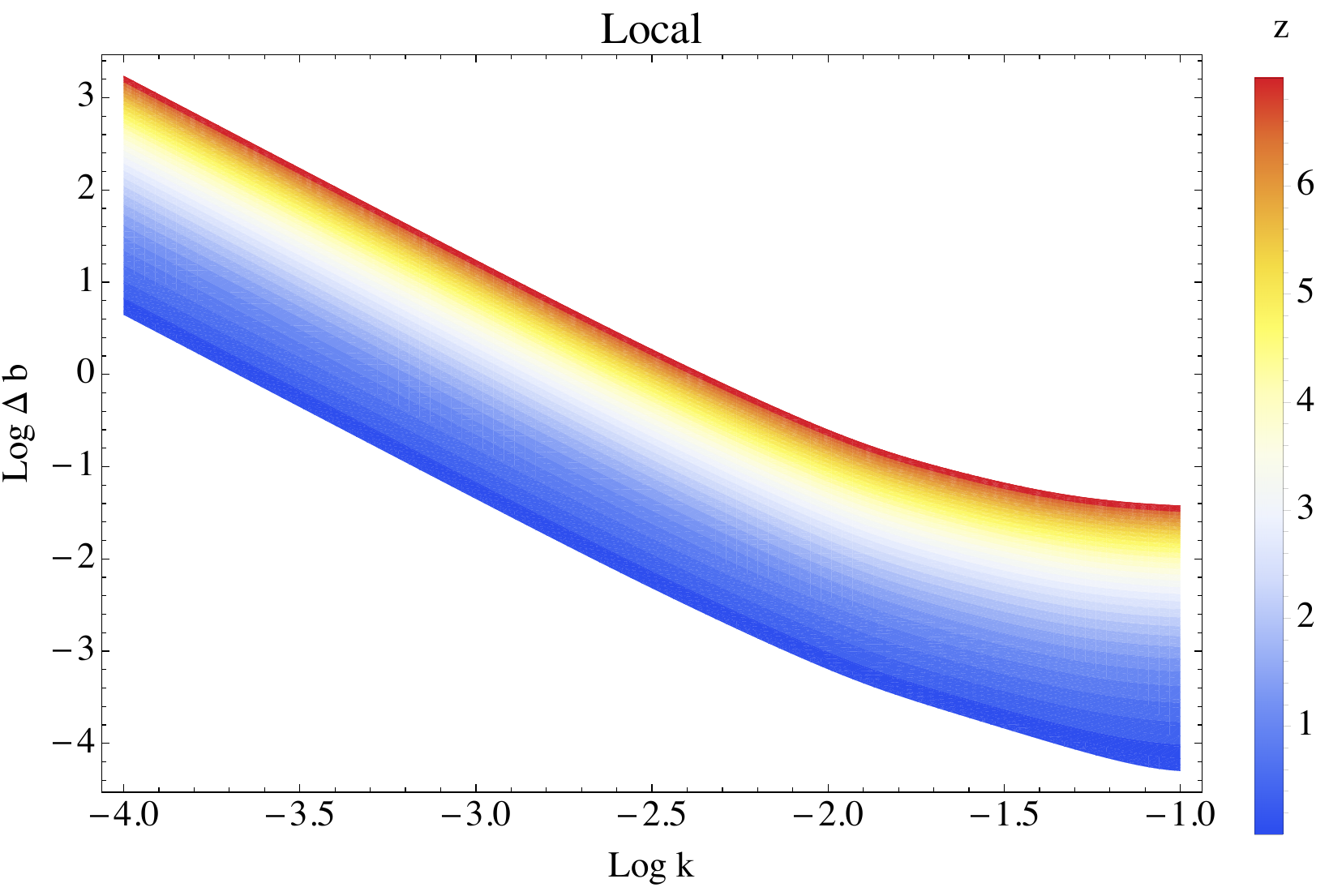}
\includegraphics[width=0.49\linewidth]{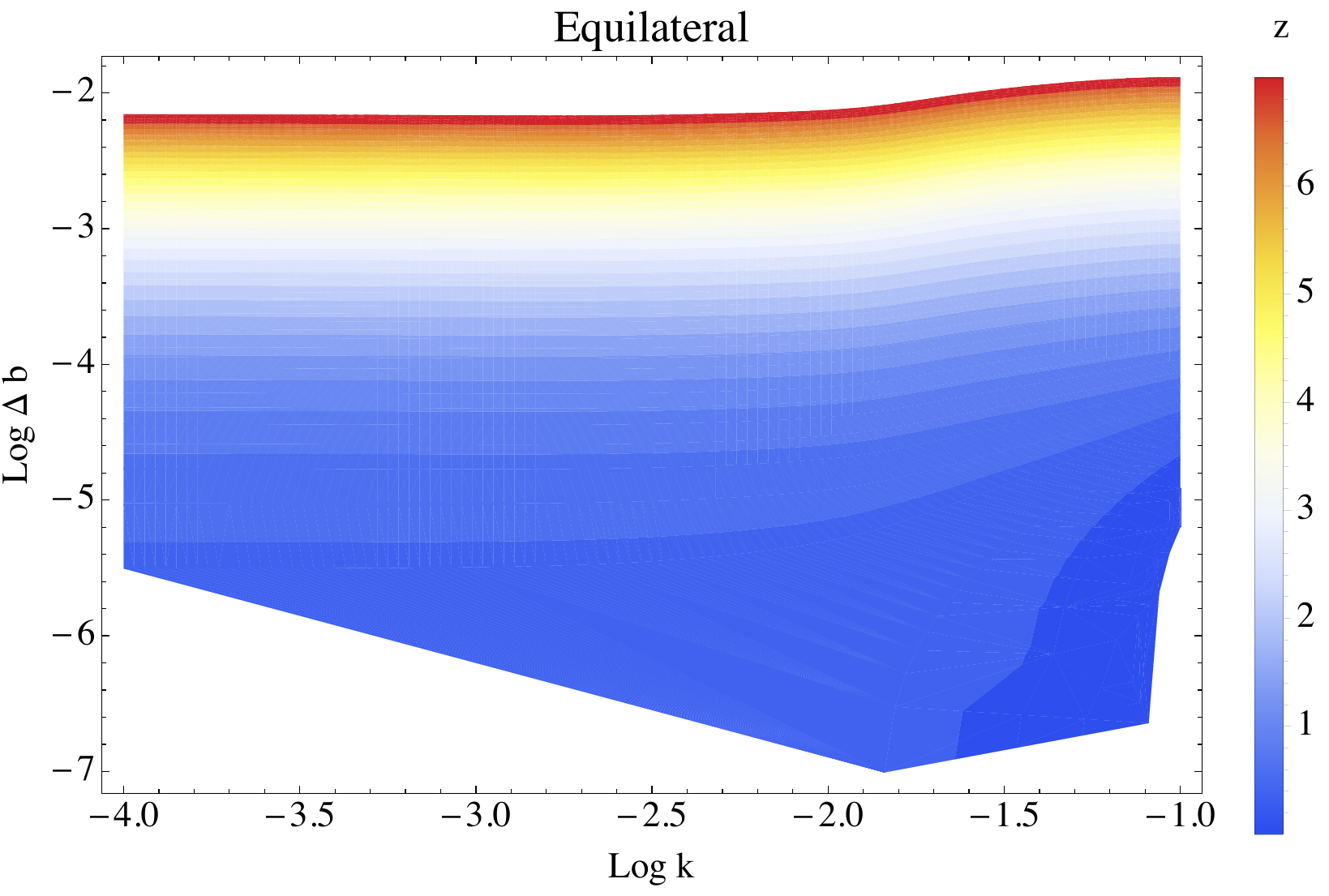}
\includegraphics[width=0.49\linewidth]{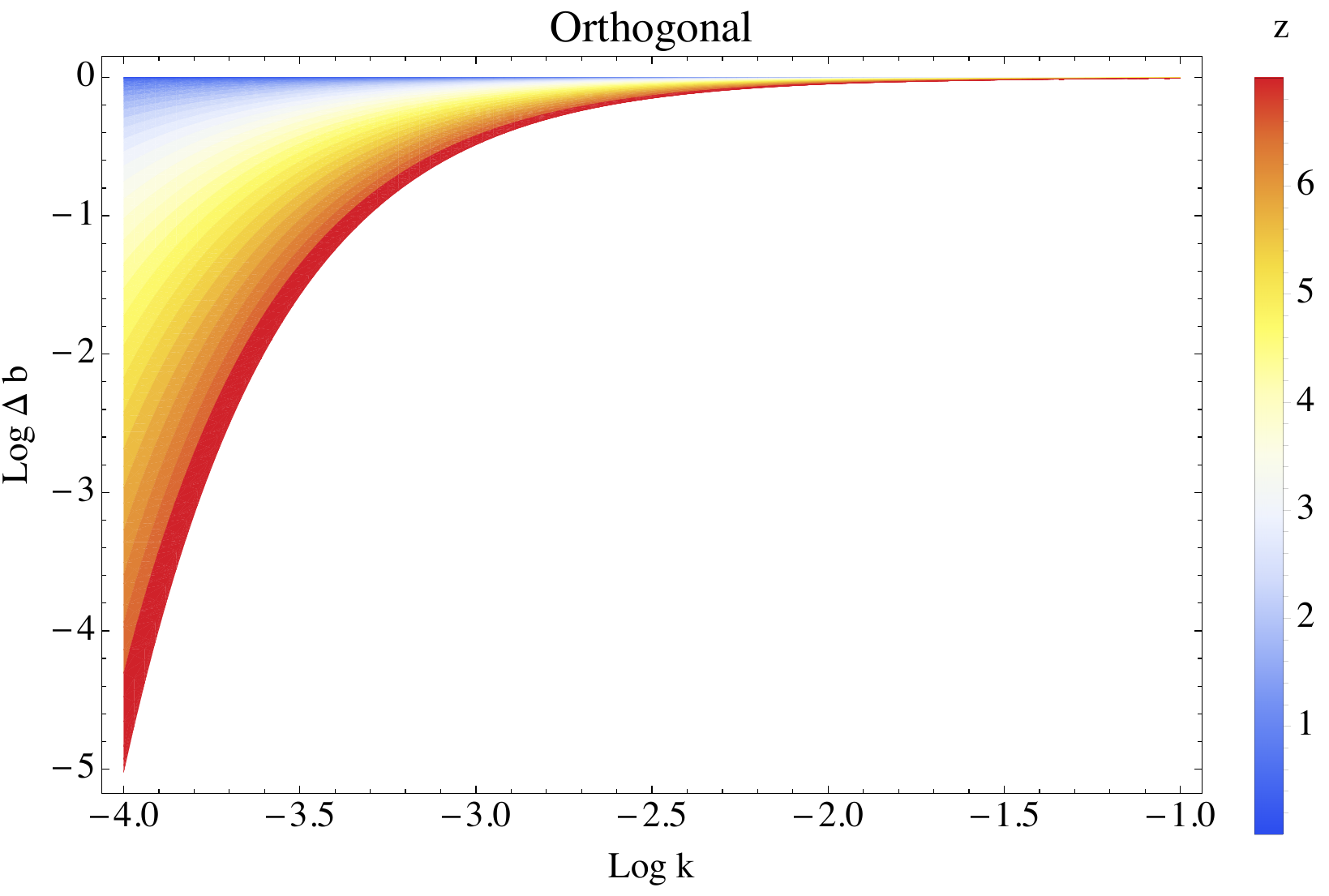}
\includegraphics[width=0.49\linewidth]{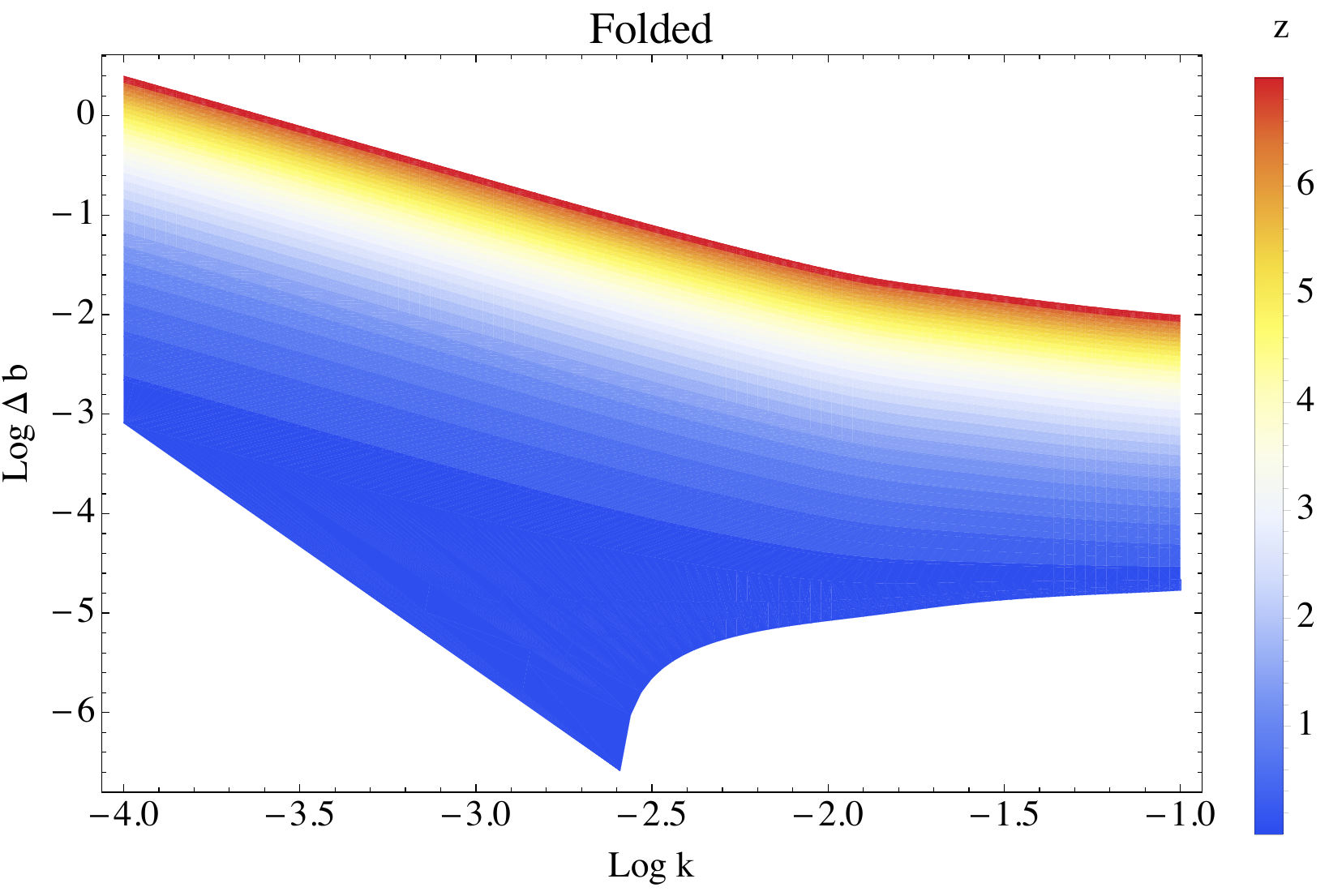}
\includegraphics[width=0.49\linewidth]{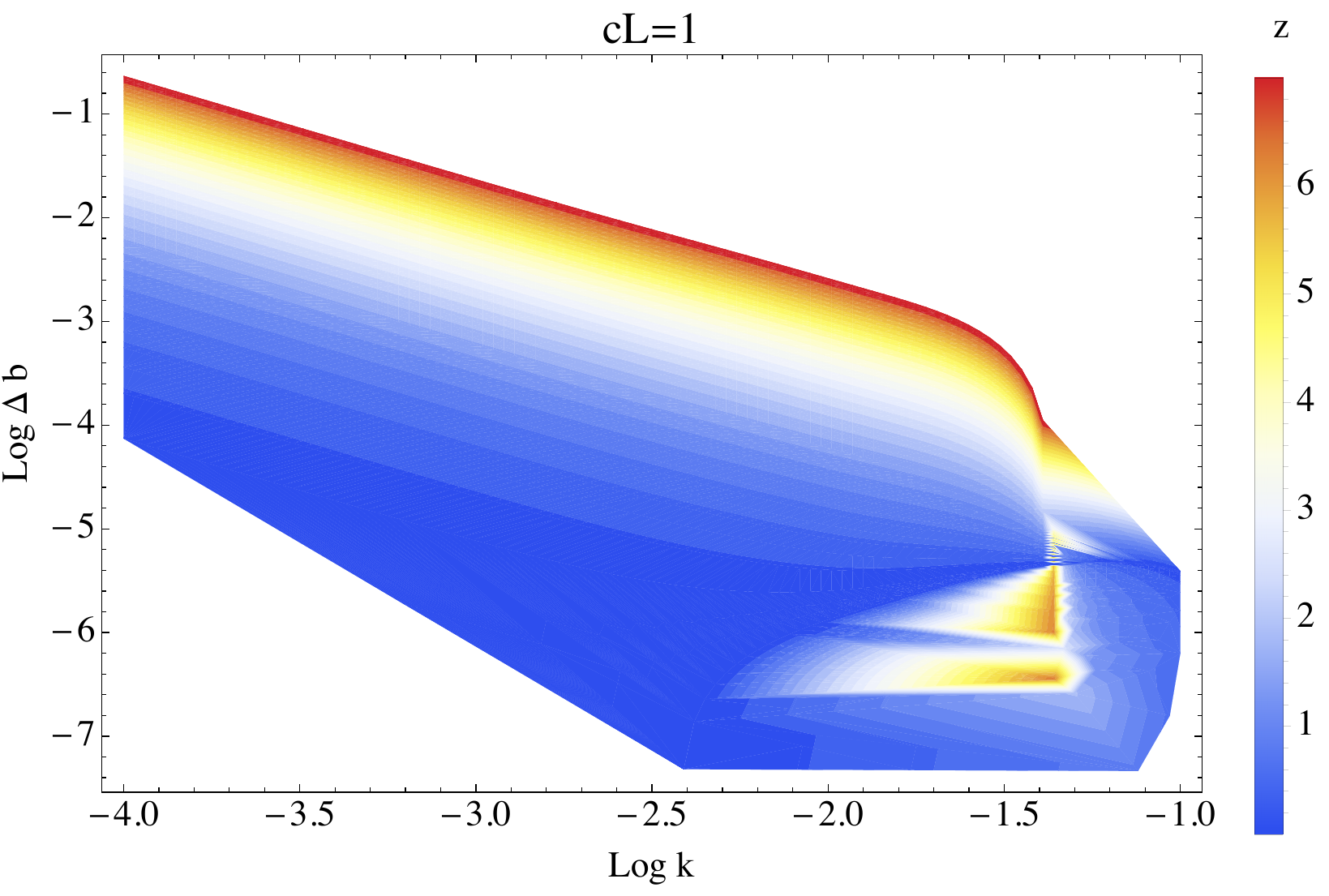}
\includegraphics[width=0.49\linewidth]{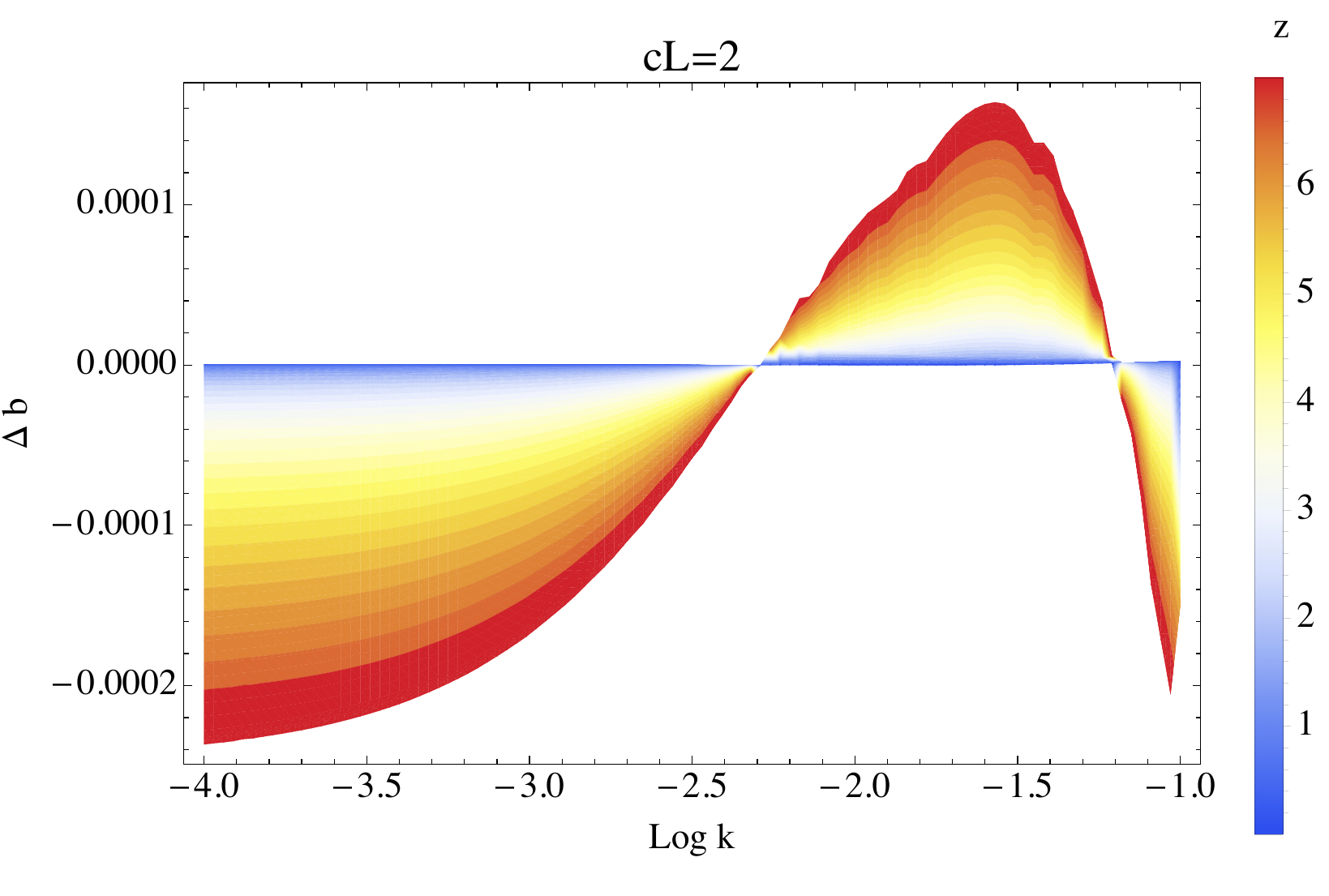}
\caption{Scale-dependent correction to the bias, for different shapes of non-Gaussianity. The relevant input parameters are fixed to the values in Figure~\ref{fig:bias_z1}. The results plotted here are for $M = 10^{14} M_\odot$.}
\label{fig:bias}
\end{figure*}

One can see the redshift dependence of $\Delta b$ for each non-Gaussianity model in Figure~\ref{fig:bias}. It is confirmed there that at higher $z$, $|\Delta b|$ grows while maintaining the overall shapes, because both $A_1$ and $A_2$ are increasing functions with redshift, on the large mass scales that give the best constraints ($M \sim 10^{14} M_\odot$)~\cite{Matsubara:2012nc}. This fact is expected to improve the sensitivity to the non-Gaussianity parameters in the analysis with high-redshift data.
In particular, we can see that e.g. in the equilateral case, the scale-dependent contribution grows with $k$ and with $z$, so deep surveys that will include both high-$z$ and high-$k$ (because of the larger $k$ for which the density field starts to be non-linear at higher redshift) measurements will sensibly improve constraints on this shape (see also~\cite{spherex}).

\section{Galaxy Surveys}
\label{sec:surveys}
We investigate constraints on various shapes of non-Gaussianity that can be set by future large-scale galaxy surveys.
In this paper we are interested in a proof-of-principle investigation to understand if future LSS surveys will allow competitive measurements of different shapes of primordial non-Gaussianity.

Measurements of non-Gaussianity have been done (and forecasted) for spectroscopic (see e.g.~\cite{Euclid, Giannantonio:2012, Ross:2013, Leistedt:2014, Byun:2014, Raccanelli:2014fNL, dePutter:2014}) and radio continuum surveys (see e.g.~\cite{Xia:2010, Raccanelliradio, Raccanelli:2015ISW, SKA:Camera}). In the first case the advantage is the high precision of the redshift information, while in the latter it is easier to survey wide areas of the sky and reach higher redshift.
Given that constraints on non-Gaussianity are in most cases better for large halo mass, scale and redshift probed, 
and that in general the effect of non-Gaussianity is larger at smaller $k$, so it is not necessary to have very precise redshift information~\cite{spherex}, we will focus on forthcoming radio surveys and assume that a combination of different surveys in optical and radio, complemented by simulations and modeling, will be able to assign redshifts to the objects observed in the planned ASKAP/EMU~\cite{EMU} and the SKA continuum~\cite{SKA:Jarvis} surveys. In both cases we assume a survey observing 30,000 deg.$^2$, with total number of objects of $\sim 7\times 10^7$ and $\sim 3\times 10^7$ for EMU (flux limit of $50 \, \mu Jy$, $100 \, \mu Jy$, respectively), $\sim 6\times 10^8$ and $\sim 5\times 10^9$ for SKA with a flux limit of $5 \, \mu Jy$ and $100 \, nJy$, respectively.

In particular, clustering information can be used to infer redshift distributions. Methods using cross-correlations of samples with unknown redshift distributions against photometric and spectroscopic datasets have been proposed~\cite{Schneider:2006, Newman:2008} and applied to real data~\cite{Ho:2008, Nikoloudakis:2012, Rahman:2014}, including the FIRST radio survey~\cite{Menard:2013}. This technique was used also to measure the redshift distribution of resolved far infrared HerMES~\cite{Mitchell-Wynne:2012} and~\cite{Schmidt:2014} cross-correlated Planck maps against quasars from SDSS to estimate the intensity redshift distribution of the Cosmic Infrared Background. A study of the impact of adding redshift information to radio continuum surveys can be found in~\cite{Camera:2012}. 

For the EMU and SKA continuum surveys we use the predicted redshift distributions, bias and halo masses of~\cite{Wilman:2008, Raccanelliradio, Norris:2012, Ferramacho:2014, SKA:Jarvis}. In particular, we assume that we will be able to distinguish different types of objects, and we follow~\cite{Ferramacho:2014} and assign halo masses as:
\begin{itemize}
\item Star forming galaxies (SFR): M = $10^{11}h^{-1} M_\odot$
\item Starbursts (SBG): M = $5\times10^{13}h^{-1} M_\odot$
\item Radio Quiet Quasars (RQQ): M = $3\times10^{12}h^{-1} M_\odot$
\item Radio loud AGN (FRI): M = $10^{13}h^{-1} M_\odot$
\item Radio loud AGN (FRII): M = $10^{14}h^{-1} M_\odot$
\end{itemize}

\begin{figure*}[htb!]
\includegraphics[width=0.49\linewidth]{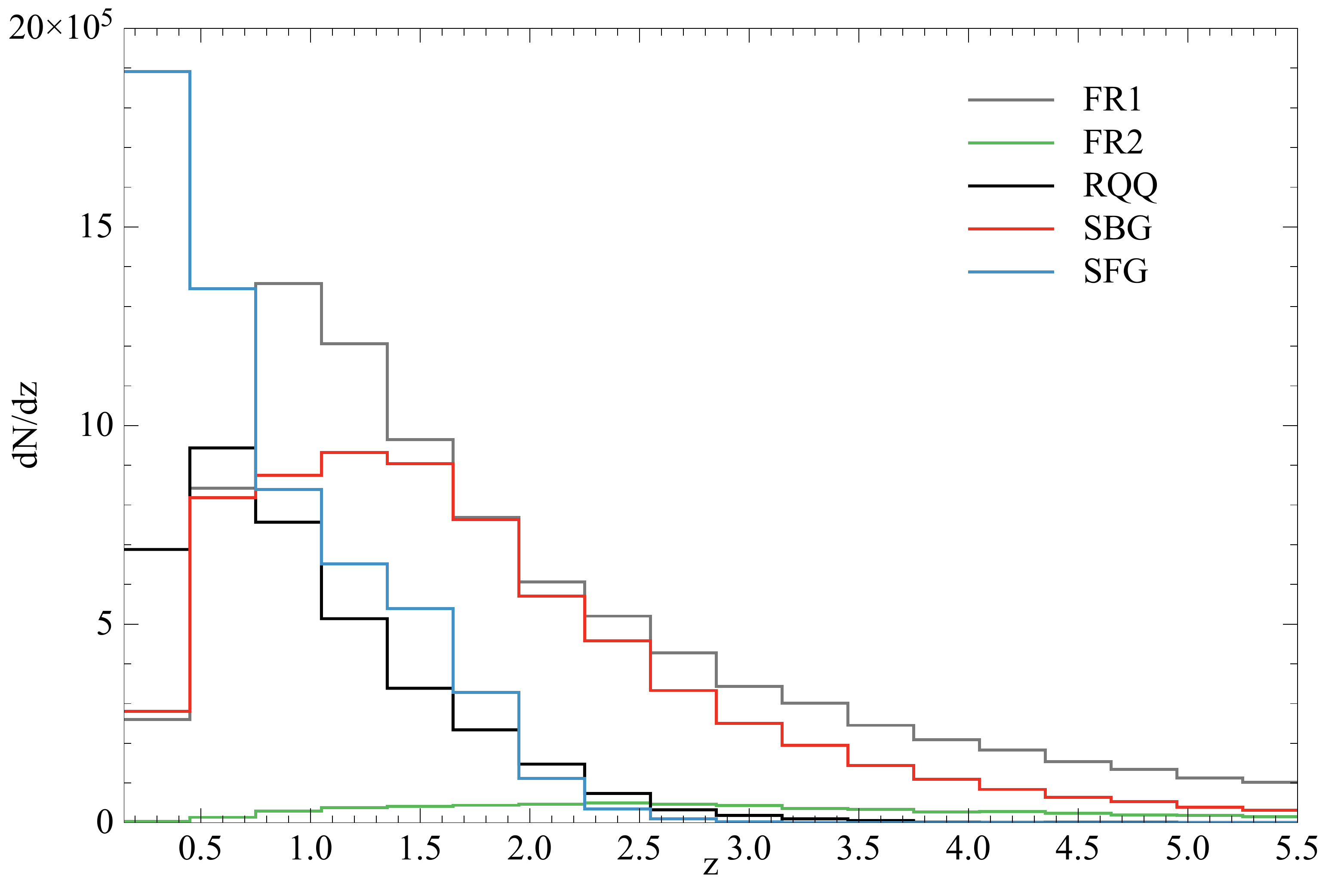}
\includegraphics[width=0.49\linewidth]{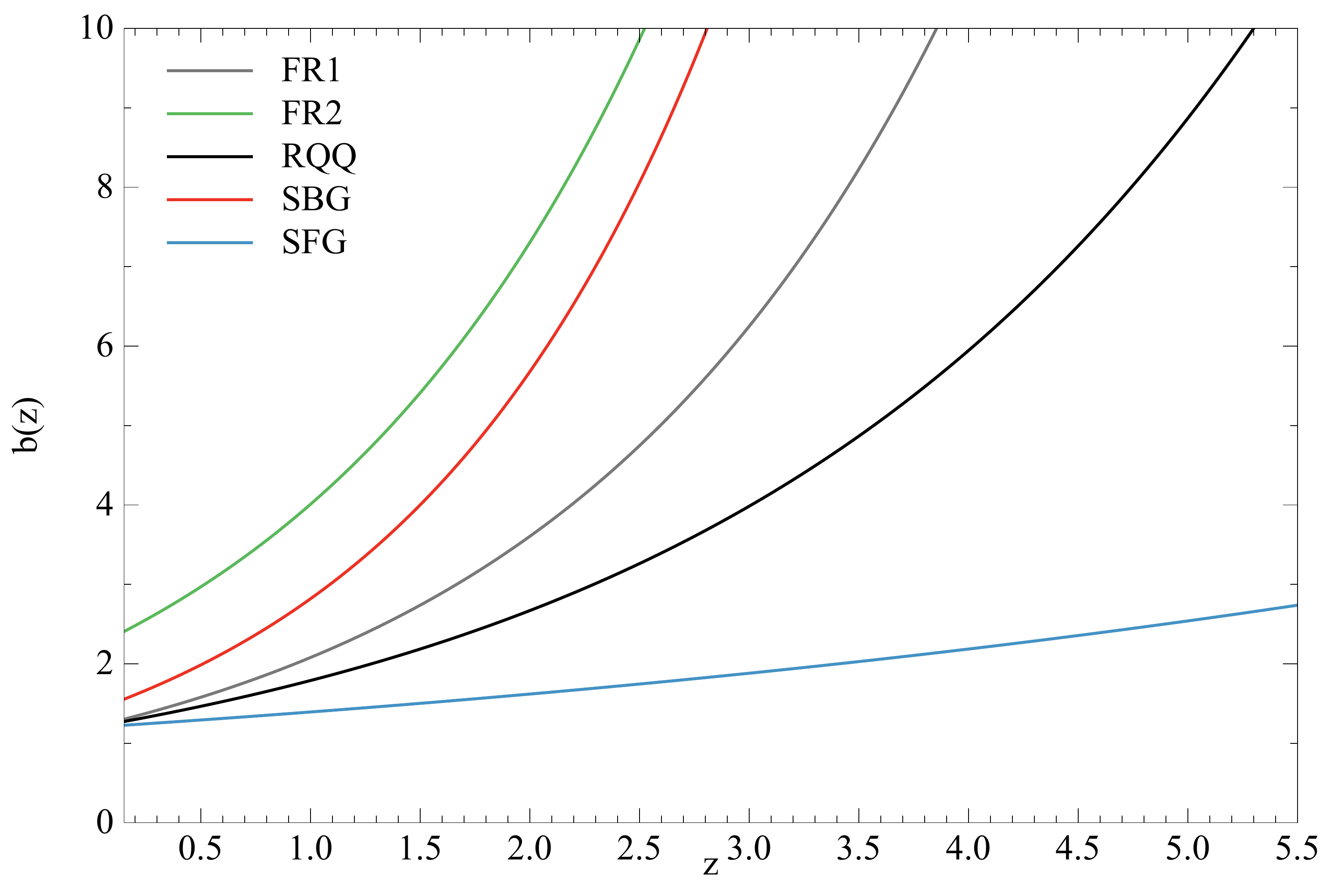}
\caption{
Redshift distribution for the ASKAP/EMU survey and bias for the different populations observed by radio surveys.
}
\label{fig:Nbz_EMU}
\end{figure*}

\begin{figure*}[htb!]
\includegraphics[width=0.49\linewidth]{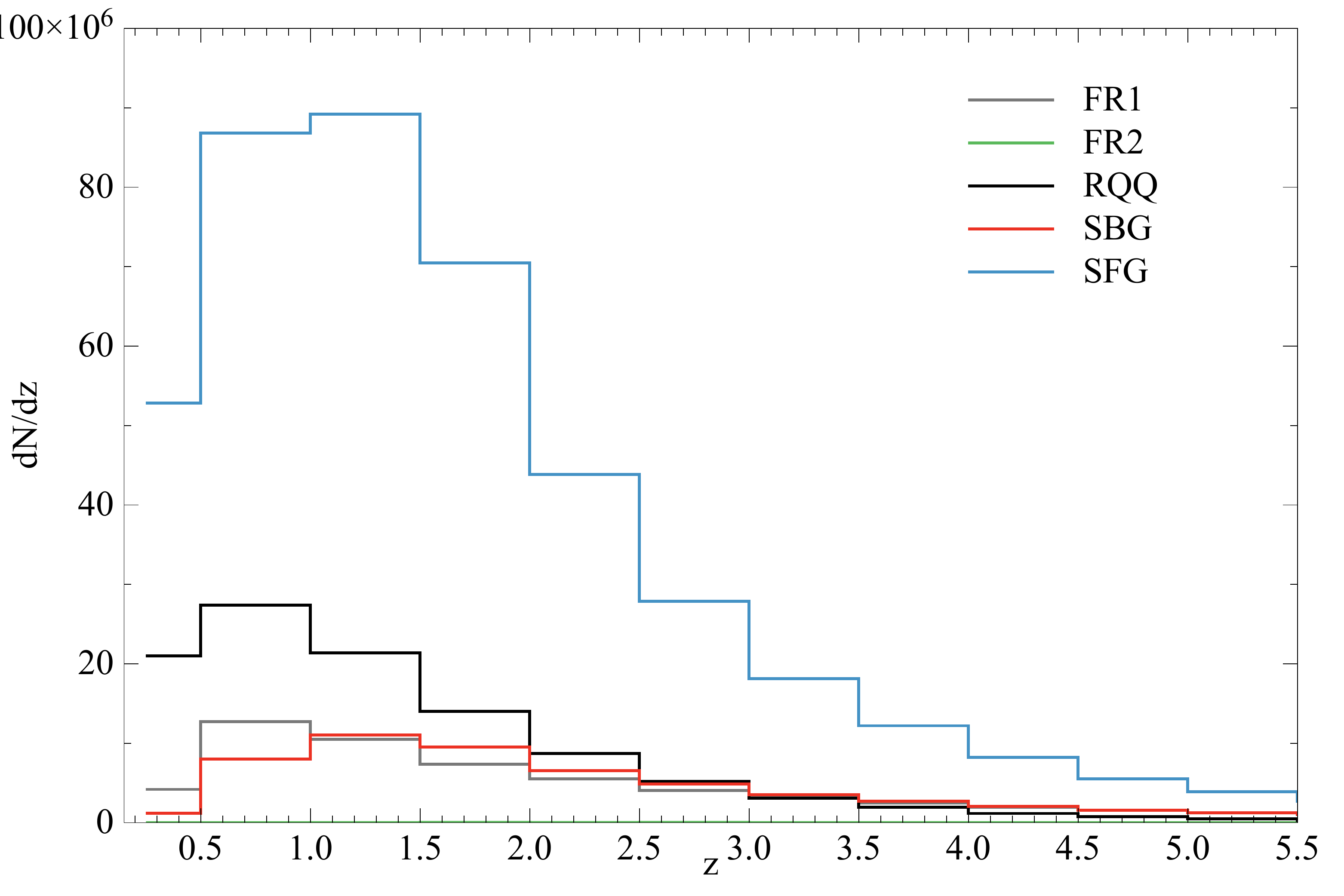}
\includegraphics[width=0.49\linewidth]{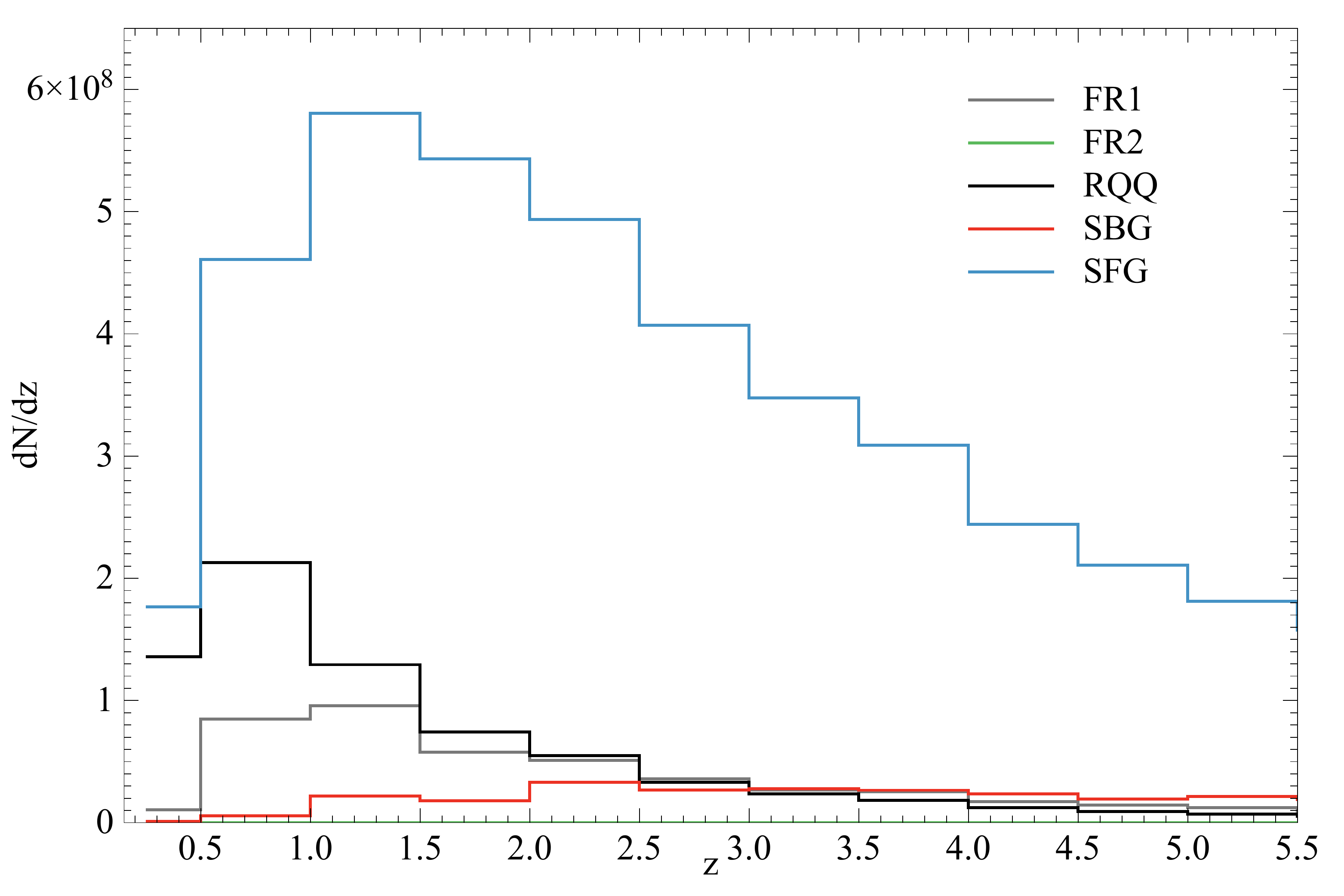}
\caption{
Redshift distributions for the different populations observed by the SKA survey with detection threshold S=5$\mu$Jy (left panel) and 100nJy (right panel).
}
\label{fig:Nz_SKA}
\end{figure*}

More details on the different cases studied are in Section~\ref{sec:results}.

\section{Power Spectrum}
\label{sec:pk}
We model the galaxy power spectrum as (see e.g.~\cite{Seo:2003}):
\begin{align}
P^s_{\rm g}(k,\mu,z) = \frac{\left[D_{\rm A}^{\rm ref}(z)\right]^2}{\left[D_{\rm A}(z)\right]^2} \frac{H(z)}{H^{\rm ref}(z)} \left[ b_{\rm NG}(k,z) + f(z) \mu^2 \right]^2 P_{\delta}^r(k,z) \, {\rm FoG}(k,\mu,z) \, {\Xi}(k,\mu,z) + P_{\rm shot}(z) \, ,
\end{align}
where the superscripts $r$ and $s$ indicate real and redshift-space, respectively, and the subscripts $\delta$ and $g$ stand for density and galaxies respectively; the superscript $^{\rm ref}$ indicates the values of the angular diameter distance and the Hubble parameter in the $\Lambda$CDM case. The shot noise contribution is the inverse of the galaxy number density at that redshift, $\left[\bar{n}_g(z)\right]^{-1}$, and the non-Gaussian bias is the one defined in Section~\ref{sec:vector}. The Redshift-Space Distortion (RSD) corrections come from the fact that the real-space position of a source in the radial direction is modified by peculiar velocities due to local overdensities; this effect can be modeled as~\cite{Kaiser:1987, Hamilton:1997}:
\begin{equation}
\label{eq:kaiser}
\delta^s(k) = \left( 1+\beta \mu^2 \right) \delta^r(k) \, ,
\end{equation}
where $\beta = f/b$, and the parameter $f$ is defined as the logarithmic derivative of the growth factor:
\begin{equation}
f = \frac{d \, \rm ln \, D}{d \, \rm ln \, a} \, .
\end{equation}
A real data analysis will need to take into account a variety of additional corrections, that will however not affect considerably our study. Examples of these corrections include observational calibrations as well as improved theoretical modeling such as wide-angle effects~\cite{Szalay:1997, Matsubara:1999, Szapudi:2004, Papai:2008, Raccanelli:2010wa,Yoo:2010,Samushia:2011,Bonvin:2011, Challinor:2011,Yoo:2012, Jeong:2012,Montanari:2012,Bertacca:2012,Raccanelli:2012growth,Raccanelli:2013,Raccanelliradial}, modifications to the observed radial redshift distribution by 
Redshift-Space Distortions~\cite{Rassat:2009}, cosmic magnification~\cite{Loverde:2007, Raccanelliradio}, and its binning~\cite{Nock:2010}; however all these are second order effects, so we will not include a detailed modeling of them.

It is also worth noting that the effect of a primordial non-Gaussianity of $\mathcal{O}(1)$ would be degenerate with large-scale effects~\cite{Bruni:2012, Maartens:2013, Raccanelli:20143D}, so in a real analysis one would need to take into account relativistic effects~\cite{Yoo:2010, Bonvin:2011, Challinor:2011,  Yoo:2012, Jeong:2012, Bertacca:2012, DiDio:2014, Camera:2014b, Alonso:2015}. Here we do not include them, because modeling them is computationally expensive, and they should not affect our work, as we focus on the precision of the measurement and not on the exact value of the non-Gaussianity parameter. A more detailed work on how GR and lensing effects can impact the observable effects of primordial non-Gaussianity shapes and if this could affect forecasts will be investigated in a future work~\cite{Bertaccaprep}. For the $f_{\rm NL}^{\rm loc}$ case, the effect of neglecting relativistic corrections on errors on $f_{\rm NL}^{\rm loc}$ and its running $n_{\rm nG}$, see~\cite{Raccanelli:2015GR}.
We will then assume for the purpose of this work that one can compute these ``relativistic corrections'' and correct for them.

Given that in this work we investigate primordial non-Gaussianity shapes that do not limit their effects on the very largest scales, it is appropriate to model more carefully the moderately large $k$ regime. We then include in our modeling of the {\it observed} power spectrum the terms FoG, $\Xi$, to model the so-called Fingers of God~\cite{Jackson:1972} and the error in the determination of the redshift of sources, respectively. We can write them as:
\begin{equation}
{\rm FoG} = e^{ -\frac{k^2 \mu^2 \sigma_v^2}{H_{\rm 0}^2} } \, , \,\,\,\,\, 
\sigma_v^2 = \frac{f^2 H_{\rm 0}^2}{6\pi^2} \int P_{\theta\theta}(k) \, dk \, ;
\end{equation}
\begin{equation}
{\rm \Xi} = e^{ - k^2 \mu^2 \sigma_z^2 } \, , \,\,\,\,\, 
\sigma_z = \frac{(1+z) \tilde{\sigma}_z}{H(z)} \, ,
\end{equation}
where $\sigma_v$ is the velocity dispersion, $P_{\theta\theta}(k)$ is the velocity power spectrum and $\tilde{\sigma}_z$ is the expected error in the determination of $z$.


\section{Results}
\label{sec:results}
In this Section we present our results. We investigate the forecasted constraints on $f_{\rm NL}$ for different shapes for future radio continuum surveys, assuming some knowledge of the redshift distribution of its sources from cross-ID and simulations, and a general spectroscopic survey.
In the first case, we also assume that we will be able to identify the type of object observed, and this will allow the implementation of the so-called multi-tracer technique~\cite{McDonald:2009, Seljak:2009}. 
Given that we want to focus on future constraints on anisotropic models, we also study how what configuration of future surveys could improve constraints on those models, and in particular what will be required to reach $\sigma (f_{\rm NL}^{c_{L=1}}) \approx 10$ or better,  which is comparable to typical values predicted by the models \cite{Shiraishi:2013vja,Bartolo:2015dga}.

Given the specifications of a survey, the Fisher matrix analysis allows us to estimate the errors on the cosmological parameters around the fiducial values (see e.g.~\cite{Fisher, Tegmark:1997, Abramo:2012}). We write the Fisher Matrix for the power spectrum in the following way:
\begin{align}
\label{eq:FM}
F_{\alpha\beta} = \int_{z_{\rm min}}^{z_{\rm max}} dz \int_{k_{\rm min}}^{k_{\rm max}}dk  \int_{-1}^{+1}d\mu
& \frac{V_{\rm eff}(k,\mu,z) k^2}{8\pi^2 \left[P(k,\mu,z)\right]^2} \frac{\partial P(k,\mu,z)}{\partial \vartheta_\alpha}\frac{\partial P(k,\mu,z)}{\partial \vartheta_\beta} B_{\rm nl} \, ,
\end{align}
where $\vartheta_{\alpha(\beta)}$ is the $\alpha(\beta)$-th cosmological parameter, and the effective volume of the survey in the z-th redshift bin is defined as:
\begin{equation}
V_{\rm eff}(k,\mu,z) = \left[\frac{\bar{n}_g(z) P(k,\mu,z)}{1+\bar{n}_g(z) P(k,\mu,z)}\right]^2 \, ;
\end{equation}
$V_s$ is the volume of the survey and $\bar{n}_g$ is the mean comoving number density of galaxies.
The last term in Eq.~(\ref{eq:FM}) accounts for non-linearities induced by the BAO peak~\cite{Seo:2003}:
\begin{equation}
B_{\rm nl} = e^{-k^2\Sigma_{\perp}^2 -k^2 \mu^2 \left( \Sigma_{||}^2 -\Sigma_{\perp}^2 \right) } ,
\end{equation}
and $\Sigma_\bot=\Sigma_0D$, $\Sigma_{||}=\Sigma_0(1+f)D$, where $\Sigma_0$ is a constant
phenomenologically describing the nonlinear diffusion of the BAO peak due to nonlinear evolution. From N-body simulations its numerical value is 12.4 $\rm h^{-1} Mpc$ and depends weakly on $k$ and cosmological parameters~\cite{Eisenstein:2007, Padmanabhan:2012}.

Eq.~(\ref{eq:FM}) involves an integral over the wavenumber $k$; for the maximum scale used it depends on the redshift bin and on the geometry of the survey, $k_{\rm min} = 2\pi V^{-1/3}$, while for $k_{\rm max}$, following~\cite{Giannantonio:2012} we use the value that gives a value of the variance $\sigma^2=0.36$.
To be completely correct, the value of $k_{\rm min}$ should be computed in a more complicated way that includes detailed considerations on the survey geometry, the radial depth of the redshift bin and large-scale modeling (see~\cite{Krause}); however, we find that a more complicated modeling of it does not considerably affects our results. We will in any case investigate below the effect of varying the choices for the $k$ range used on the measurements of $f_{\rm NL}$ for the anisotropic models.

We present our results for a Fisher analysis focusing only on the $f_{\rm NL}$ parameter, assuming knowledge of cosmological parameters describing e.g. dark energy and curvature, and we allow an overall amplitude of the gaussian bias $b_G$ (that might include  other amplitude uncertainties) to vary. Real constraints might be affected by the uncertainties on other cosmological parameters, but here we focus on errors on the scale-dependent bias, and we argue that future priors from CMB and other LSS experiments will mitigate the effect on larger Fisher matrices.

We make use of the so-called multi-tracer technique, that allows the reduction of cosmic variance errors for surveys that are observing (and able to discriminate) different types of objects, in particular objects having different bias values. This technique was originally proposed in~\cite{McDonald:2009, Seljak:2009} and then further studied in e.g.~\cite{Hamaus:2010qr}; in~\cite{Hamaus:2011fp} it was in particular applied to non-Gaussianity analyses. A real data application was performed in~\cite{Blake:2013}.

A detailed analysis of how much one can gain using more than one tracer depends on the details of the observed galaxy catalog; in particular, the ideal case (see e.g.~\cite{Seljak:2009}) is when a survey can target a very large number of unbiased sources and a subsample of highly biased objects (so a practical way to obtain a more efficient multi-tracer analysis could then be by combining a spectroscopic with a photometric catalog). An investigation of the multi-tracer technique for designing a survey measuring the local non-Gaussianity parameter $f_{\rm NL}^{\rm loc}$ was recently presented in~\cite{dePutter:2014}.
For the practical implementation of this technique we follow~\cite{Blake:2013} and we use the 5 populations described in Section~\ref{sec:pk}.
In this case the Fisher matrix becomes:
\begin{equation}
F_{\alpha\beta} = \int dz \, dk \, d\mu \sum_{i,j} \Upsilon(k,\mu,z) \frac{\partial P_{ij}(k,\mu,z)}{\partial \vartheta_{\alpha}} \left[\mathcal{C}(k,\mu,z)^{-1}\right]_{ij} \frac{\partial P_{ij}(k,\mu,z)}{\partial \vartheta_{\beta}} \, ,
\end{equation}
where $\Upsilon(k,\mu,z)$ includes the effective volume and the other terms other than the derivatives of Eq.(~\ref{eq:FM}), $P_{ij} = b_ib_jP_\delta+\delta_{ij}/n_i$, and $\mathcal{C}$ is the covariance matrix including auto- and cross- spectra (see~\cite{Blake:2013} for details).

Here we present our findings when using the radio continuum survey as in Figure~\ref{fig:Nbz_EMU}, modeled after the ASKAP/EMU survey~\cite{EMU}, for the predicted detection threshold of S = 50$\mu$Jy, and for a more conservative case of S = 100$\mu$Jy. We then investigate the improvements in the constraining power when using an SKA-like survey, assuming a detection threshold of S = 5$\mu$Jy, and an optimistic S=100 nJy (see~\cite{SKA:Jarvis} for details).
A $P(k)$ analysis using radio continuum data will require cross-identification of sources in order to assign a redshift to the objects observed. This can be possible by the combination of data with optical surveys and the modeling of the redshift distribution of sources using simulations.
We compute constraints on the different shapes of non-Gaussianity when assuming an EMU survey such as the one of~\cite{Raccanelliradio, Norris:2012}, and the two SKA cases, considering three different assumptions for the redshift information that will be available for the radio sources detected. We consider:
\begin{itemize}
\item a Conservative case, where we divide the catalog into 3 redshift bins, $0<z<1$, $1<z<2$, $2<z<6$, with $\sigma(z)=0.1$;
\item a Optimistic case, where we divide the catalog into 5 redshift bins, $0<z<0.5$, $0.5<z<1$, $1<z<2$, $2<z<3$, $3<z<6$, with $\sigma(z)=0.03$;
\item a Futuristic case, where we divide the catalog in bins of $\Delta z =0.5$ in the range $0<z<6$, with $\sigma(z)=0.001$.
\end{itemize}
In Table~\ref{tab:results_emuska} we show our results for the above cases. It can be seen that the quality of the redshift information available makes a huge difference.

Errors on the vector fields coefficients will be competitive with CMB ones only for the optimistic SKA 5$\mu$Jy case, or a survey comparable to SKA 100 nJy will be needed. For the other shapes, we can see that galaxy surveys can more easily be competitive.

In particular, the EMU survey for the conservative case would not be competitive in constraining non-Gaussianity shapes; in the optimistic case, on the other hand, constraints would be comparable to constraints coming from the CMB for the 100$\mu$Jy survey. The 50$\mu$Jy case, instead, would already represent an improvement in measuring $f_{\rm NL}$.
Having a result comparable to the CMB one would already be a very good result, as different measurements on different wavelengths, redshift, and instruments would represent an important check of {\it Planck} results.
It is interesting to note that, in the 5 bins case, going from S=100 to 50 $\mu$Jy would allow the EMU survey to surpass Planck in the precision of $f_{\rm NL}$ measurements.

For the SKA surveys, even in the conservative case with the 5$\mu$Jy detection limit, results are already very interesting and show how LSS can realistically improve upon CMB constraints in the not so far future. In the more advanced case of the SKA with a 100 nJy detection limit, predicted constraints are impressive and they would allow very stringent tests of inflationary models.
The futuristic case analyzed here is not necessarily related to radio continuum surveys, but it is a sort of lower limit for the errors that this type of analysis could obtain with a future survey. In this case, we model the survey specifications after the SKA survey with S=100 nJy, but with redshift information precise enough to divide the catalog in bins of $\Delta z = 0.5$.
It represent more of a proof-of-principle investigation, and results are intended to show that in principle LSS can allow extremely precise tests of non-Gaussianity of any shape.
We also note that our predictions, considering different approaches in the analyses, are in good agreement with predictions for the $f_{\rm NL}^{\rm loc}$ model of e.g.~\cite{Raccanelli:2015ISW, Camera:2014a, Camera:2014b, Ferramacho:2014, SKA:Camera}.

\begin{table}[h]
{\small
\begin{tabular}{|l|cc|cc|cc|cc|c|c|}
\hline
Shape & EMU$_C^{100\mu Jy}$ & EMU$_O^{100\mu Jy}$ & EMU$_C^{50\mu Jy}$ & EMU$_O^{50\mu Jy}$ & SKA$_{C}^{5\mu Jy}$ & SKA$_{O}^{5\mu Jy}$ & SKA$_{C}^{100 nJy}$ & SKA$_{O}^{100 nJy}$ & Futuristic & CMB \\ \hline
$\sigma(f_{\rm NL})$ local & 11.94 & 5.54 & 9.26 & 4.37 & 1.62 & 1.06 &  0.67& 0.51 & 0.21 & 5.7 \\
$\sigma(f_{\rm NL})$ equilateral & 221.14 & 79.84 & 179.03 & 62.58 & 22.24 & 9.09 & 6.18 & 2.83 & 0.42 & 70 \\
$\sigma(f_{\rm NL})$ orthogonal & 102.97 & 39.04 & 82.25 & 30.69 & 15.30 & 7.40 & 6.71 & 3.35 & 0.54 & 33 \\
$\sigma(f_{\rm NL})$ folded & 151.48 & 56.45 & 121.50 & 44.35 & 20.29 & 9.25 & 8.15 & 4.14 & 0.96 & 65 \\
$\sigma(c_{L=1})$ & 1916.29 & 721.15 & 1519.8 & 558.35 & 200.62 & 78.32 & 41.81 & 17.54 & 2.14 & 103 \\
$\sigma(c_{L=2})$ & 10874.9 & 4113.82 & 8436.7 & 2952.5 & 1098.93 & 393.60 & 193.48 & 76.55 & 9.22 & 26 \\
\hline
\end{tabular}
}
\caption{Results for predicted EMU (with detection threshold S=100$\mu$Jy and 50$\mu$Jy) and SKA-like surveys (with S=5$\mu$Jy and 100 nJy), assuming the distribution of sources and their bias as in Figures~\ref{fig:Nbz_EMU}, ~\ref{fig:Nz_SKA}.
Results are for the conservative (C) and optimistic (O) cases. We also compare with a futuristic survey, representing an upper limit for constraints from this type of analysis. See text for more details. For comparison we here depict the CMB temperature limits (68\%CL) obtained by the {\it Planck} collaboration \cite{plancknG2015}.
}
\label{tab:results_emuska}
\end{table}

\section{Future constraints on vector fields models}
\label{sec:vec_con}
\begin{figure*}[htb!]
\includegraphics[width=0.49\linewidth]{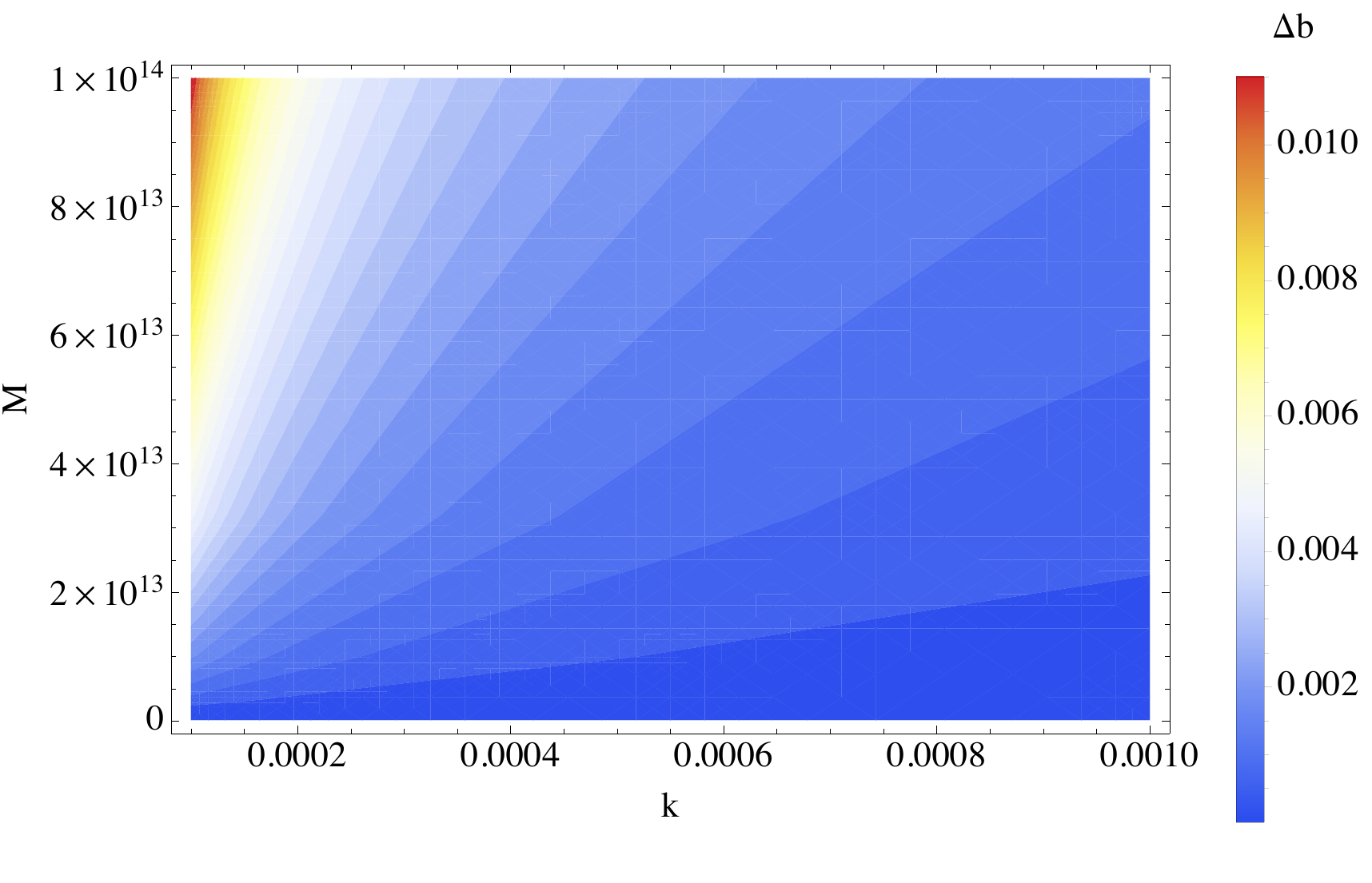}
\includegraphics[width=0.49\linewidth]{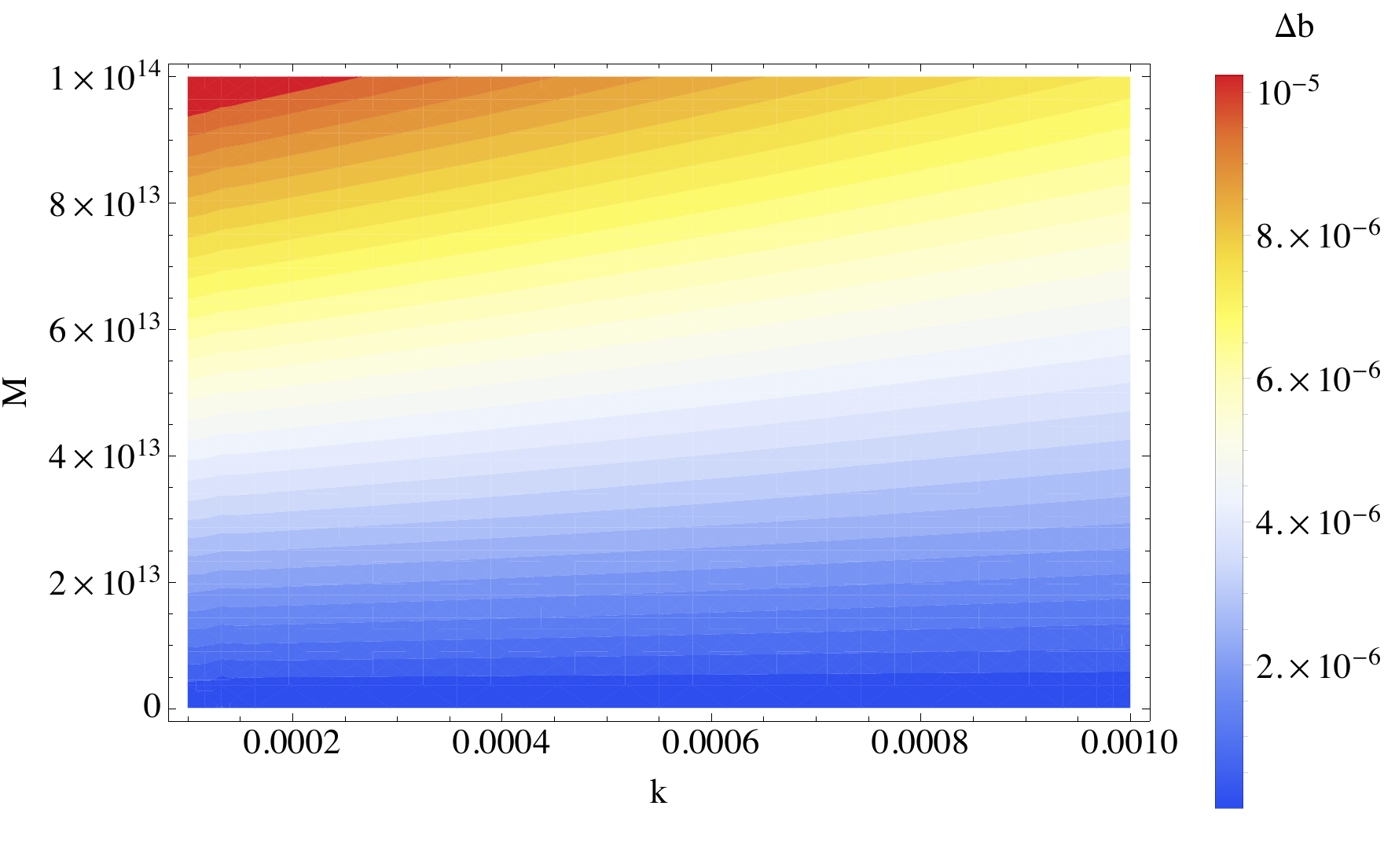}
\includegraphics[width=0.49\linewidth]{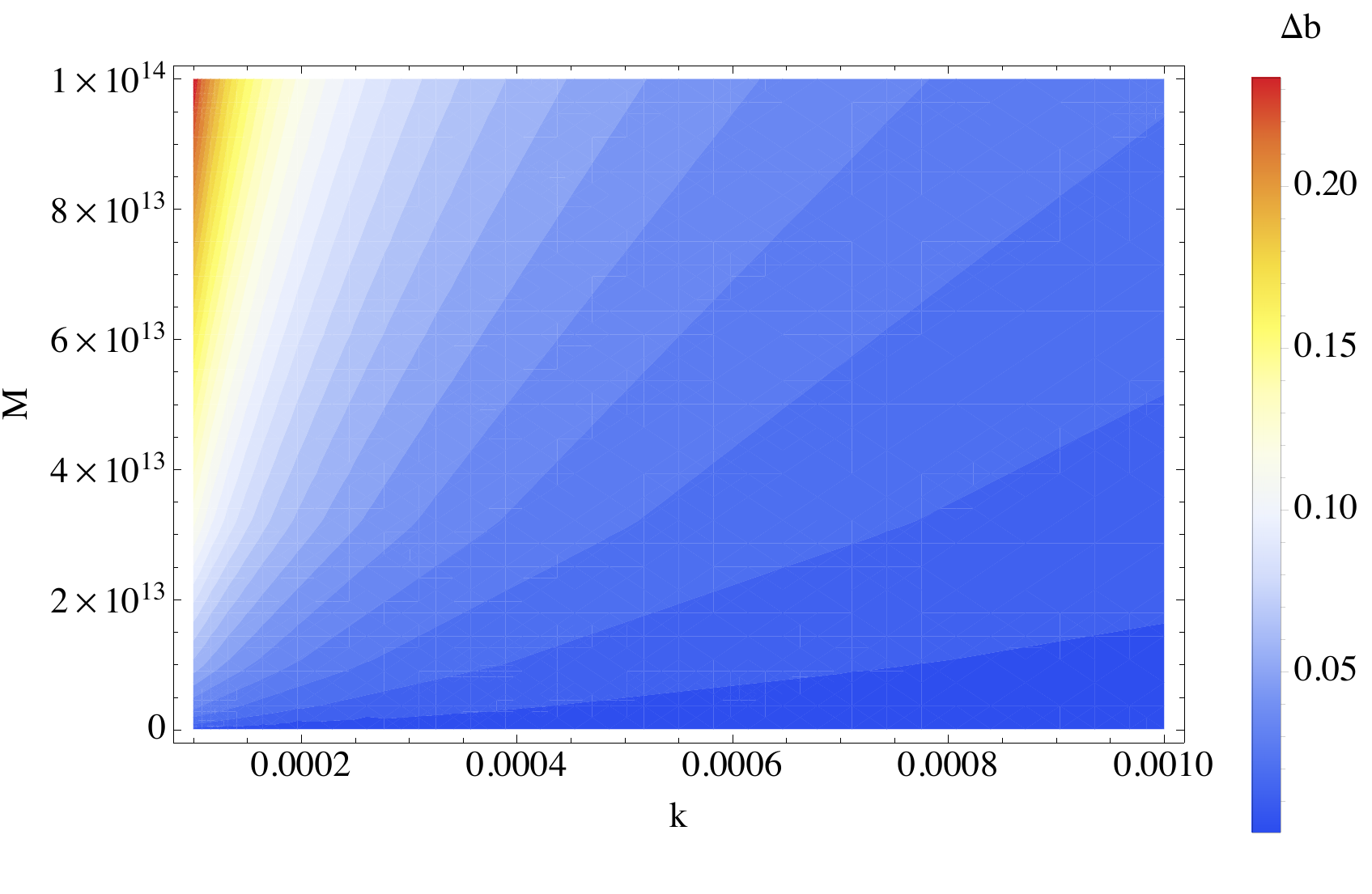}
\includegraphics[width=0.49\linewidth]{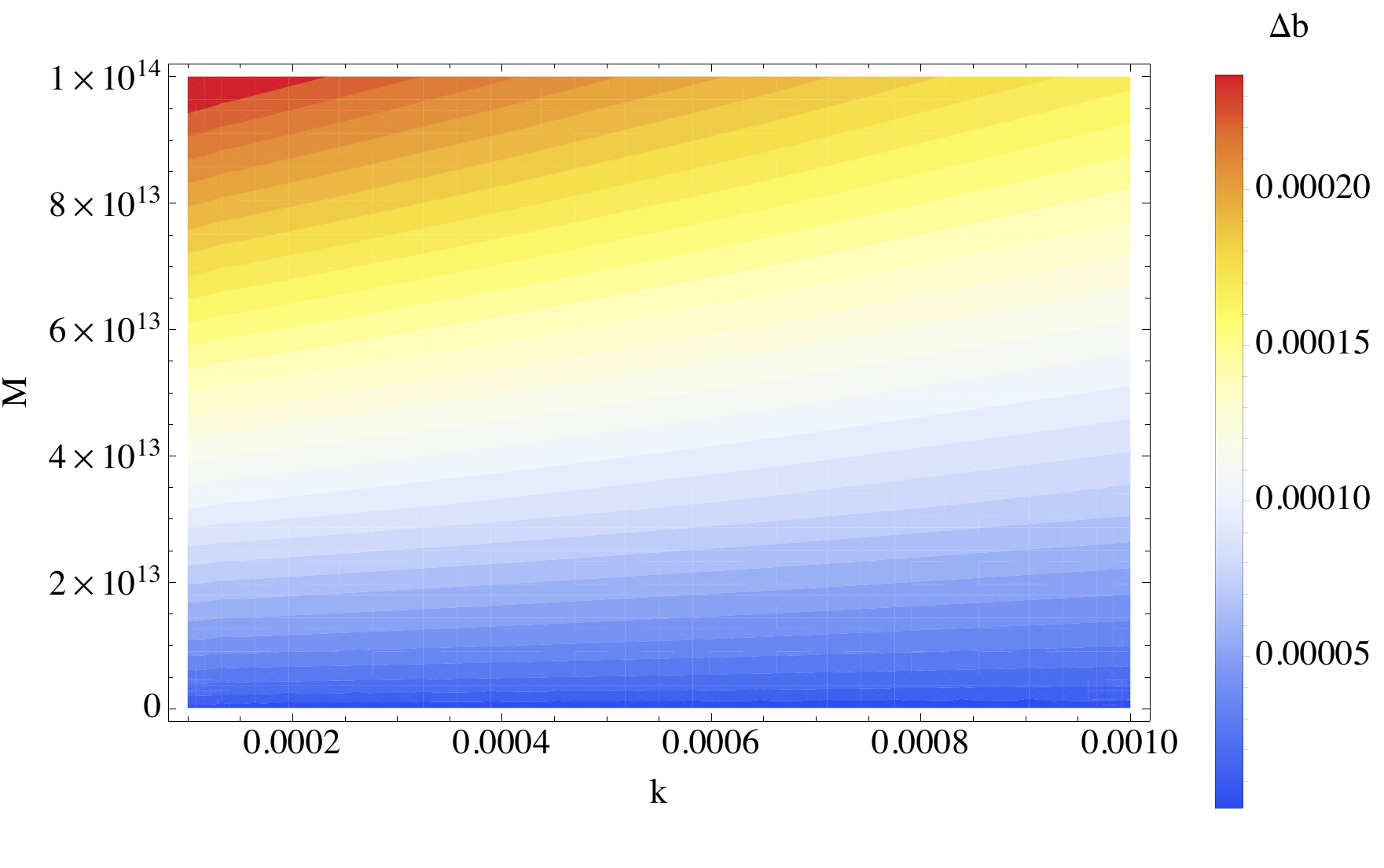}
\caption{Non-Gaussian correction to the bias as a function of scale $k$ and halo mass, for the $c_{|L=1}$ (left panels) and $c_{|L=2}$ (right panels) parameters. Top panels are for $z=2$, bottom panels for $z=7$.}
\label{fig:cLiz}
\end{figure*}

\begin{figure*}[htb!]
\includegraphics[width=0.49\linewidth]{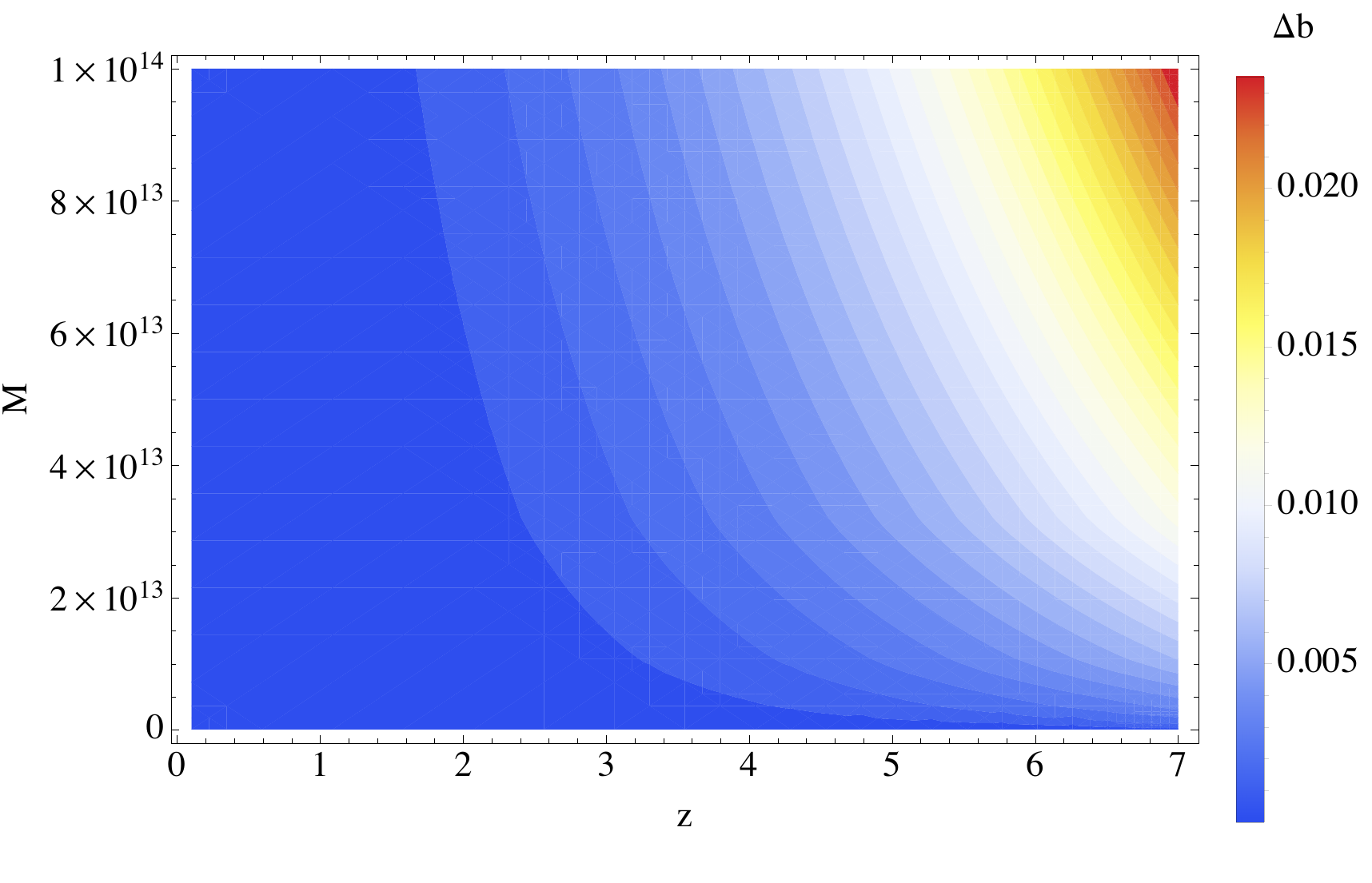}
\includegraphics[width=0.49\linewidth]{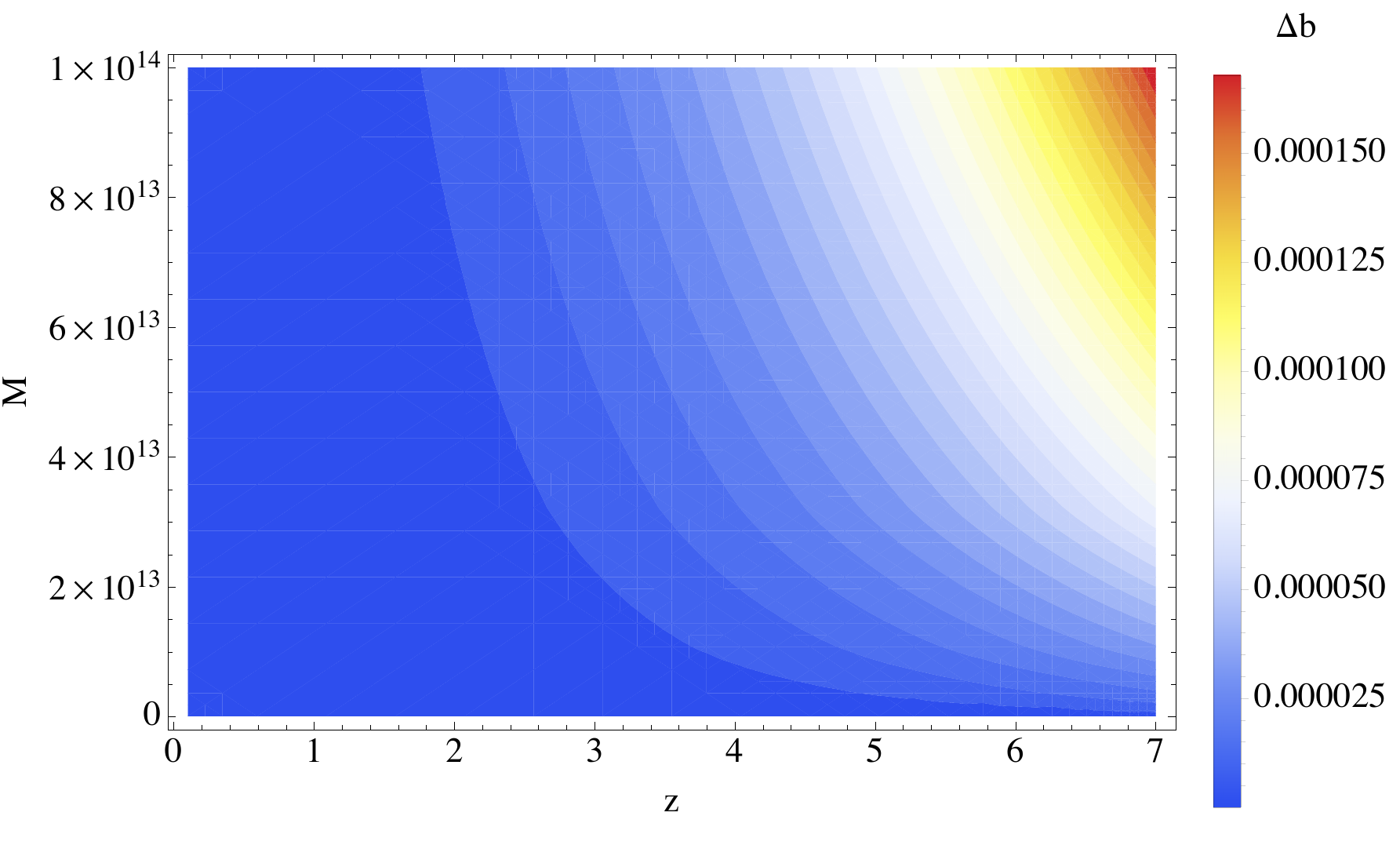}
\includegraphics[width=0.49\linewidth]{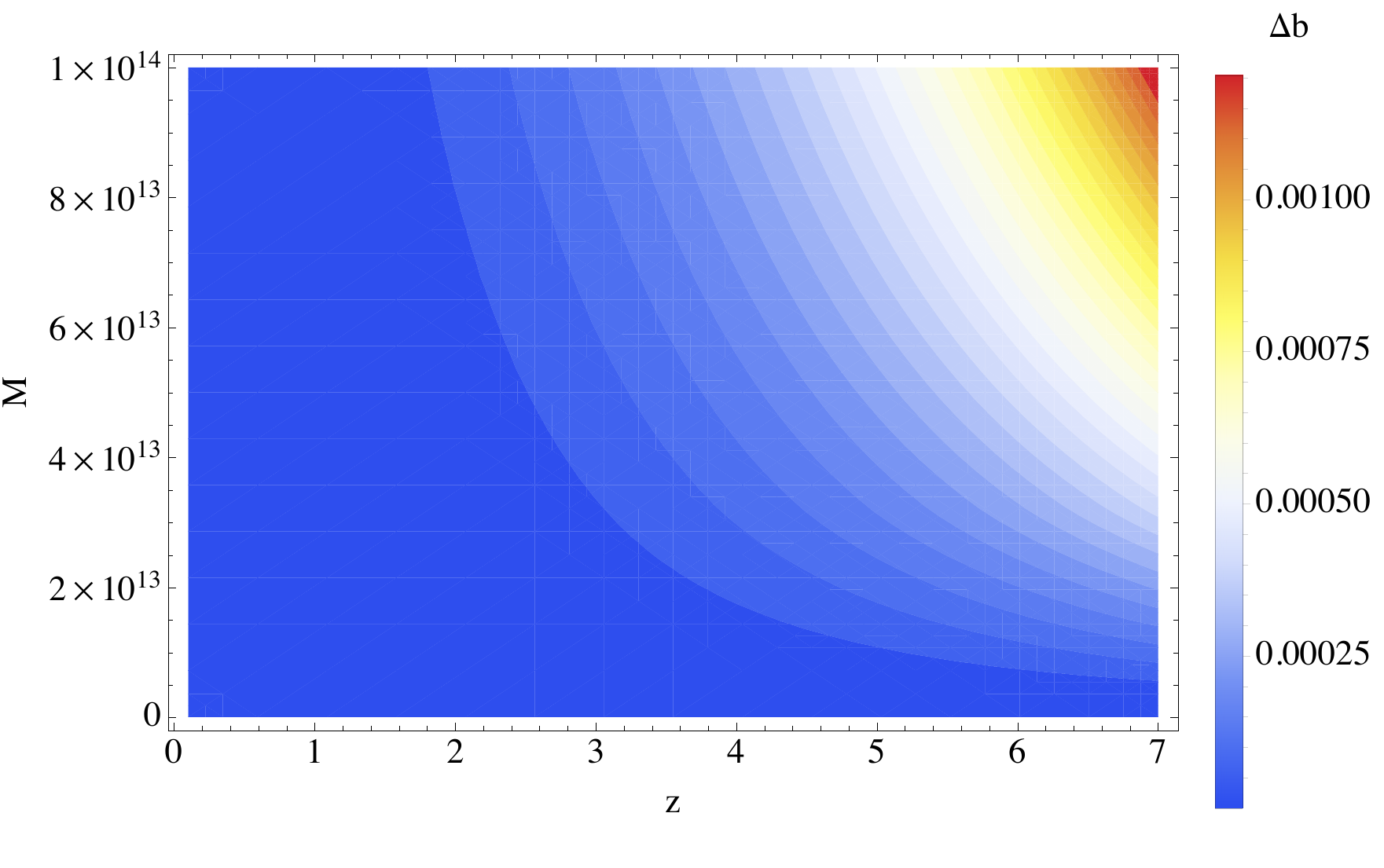}
\includegraphics[width=0.49\linewidth]{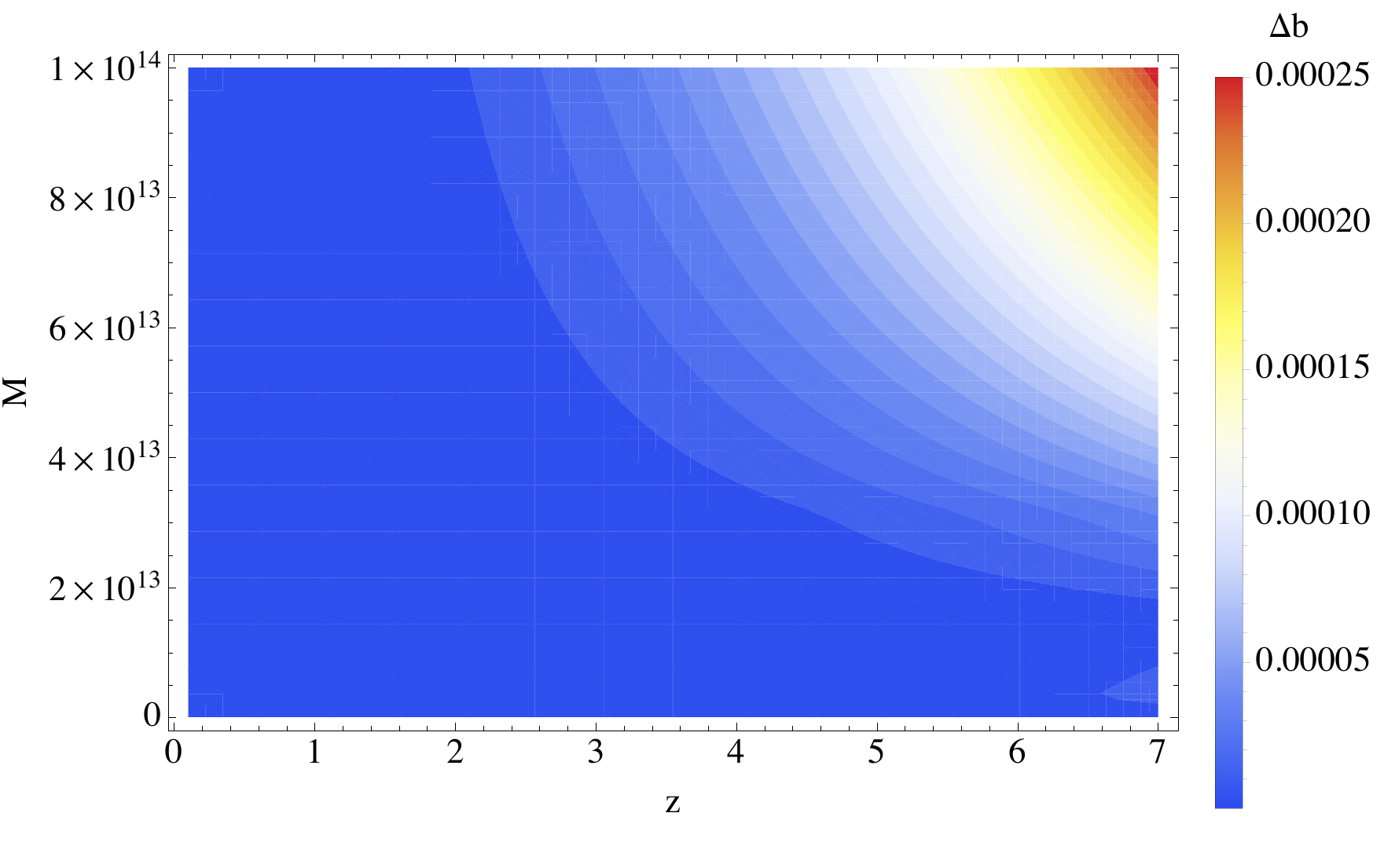}
\caption{Non-Gaussian correction to the bias as a function of redshift $z$ and halo mass, for the $c_{|L=1}$ (left panels) and $c_{|L=2}$ (right panels) parameters. Top panels are for $k=0.001$, bottom panels for $k=0.1$.}
\label{fig:cLik}
\end{figure*}

Given that the scope of this paper is mainly to investigate if LSS can set meaningful constraints on angle-dependent non-Gaussianity, in this Section we try to find the minimum requirements for a galaxy survey to set meaningful constraints on the vector fields model parameters $c_L$. In Figures~\ref{fig:cLiz},~\ref{fig:cLik} we plot the non-Gaussian correction to the bias for the $c_{|L=1,2}$ parameters as a function of redshift, scale and halo mass. It is clear that for the $c_{|L=1}$ parameter, very large scales, very high redshifts and very large halo masses are required to obtain strong constraints.
The $c_{|L=2}$ parameter, on the other hand, exhibits a different behavior; the amplitude of $\Delta b$, while decreasing still with halo mass and redshift, decays less dramatically with scale. For this reason, as we saw before, increasing the maximum $k$ used in the analysis largely increases the constraining power.

In the rest of this Section we investigate what are the minimum requirements needed for a galaxy survey in order to reach a precision on measurements of $c_{|L=1} \approx 10$, which can be considered as a typical (in some cases lower) value predicted by the scenarios that produce this kind of non-gaussian signature~\cite{Shiraishi:2013vja,Bartolo:2015dga}~\footnote{Notice that in the case of the inflationary models proposed in~\cite{Bartolo:2015dga} the natural values predicted for $c_1$ are typically larger than $10$.}; it is clear from the results in Table~\ref{tab:results_emuska} and Figures~\ref{fig:cLiz},~\ref{fig:cLik}, that it is unlikely LSS will be competitive with CMB experiments in measuring the parameter $c_{|L=2}$.

\begin{figure*}[htb!]
\includegraphics[width=0.49\linewidth]{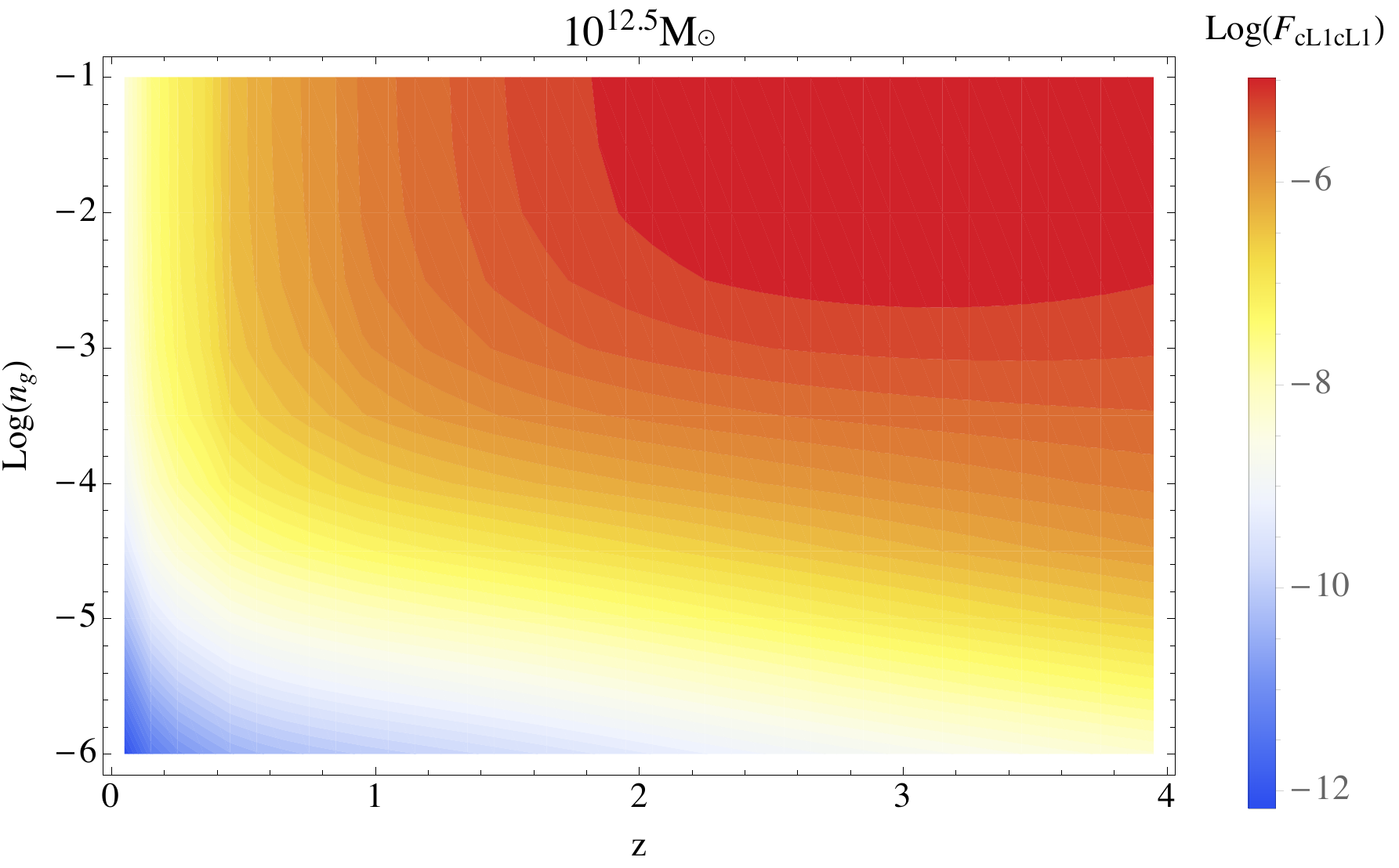}
\includegraphics[width=0.49\linewidth]{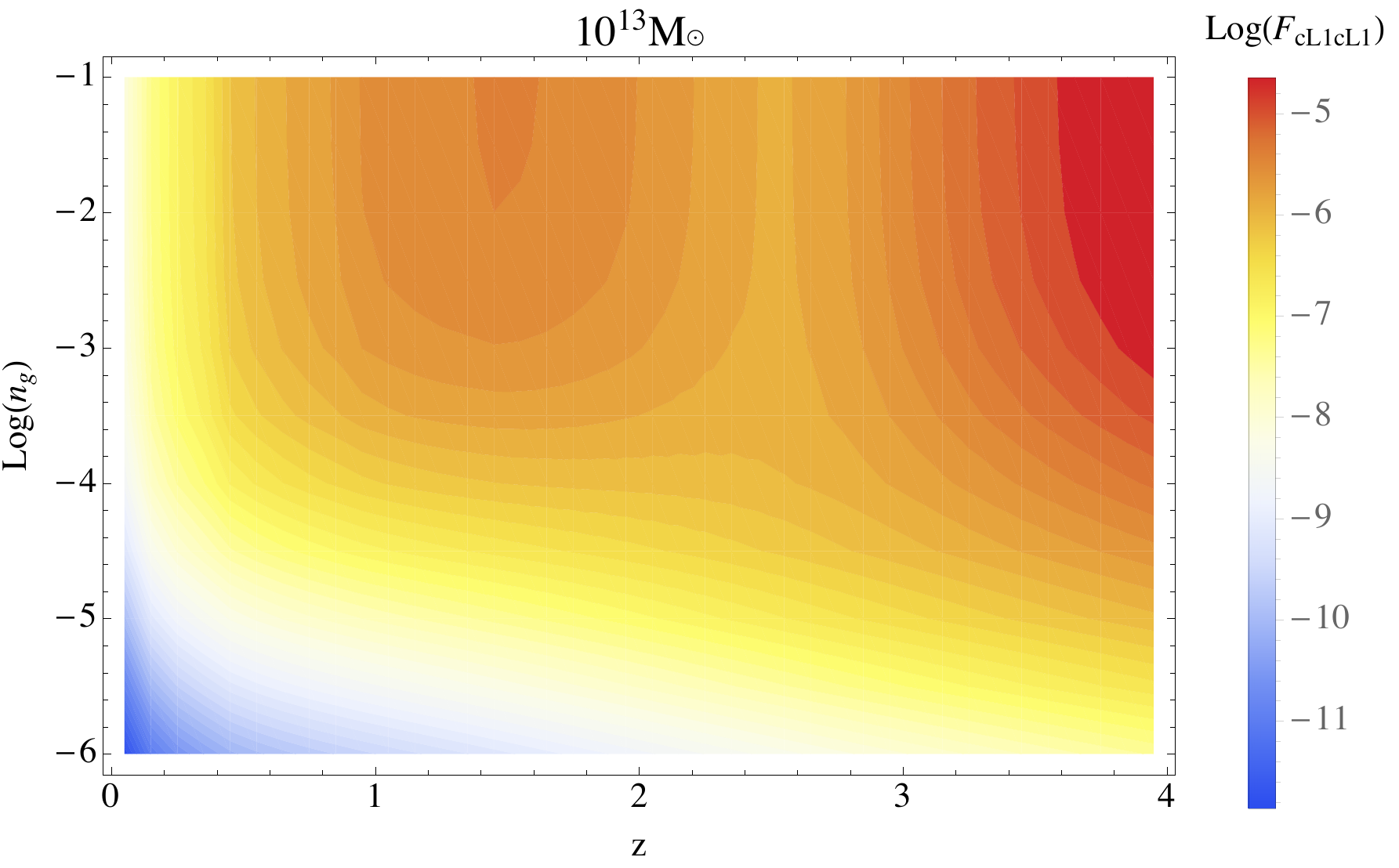}
\includegraphics[width=0.49\linewidth]{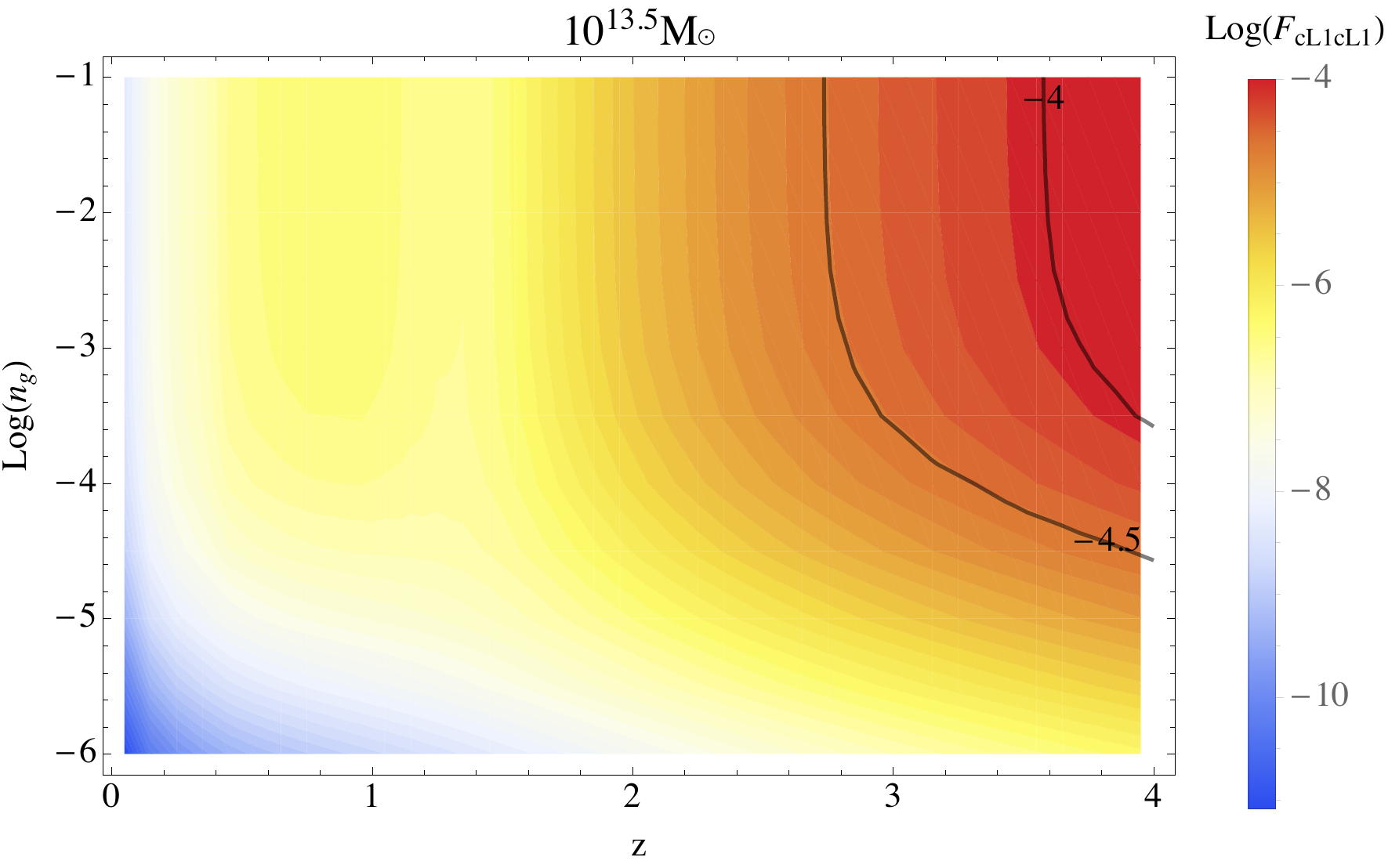}
\includegraphics[width=0.49\linewidth]{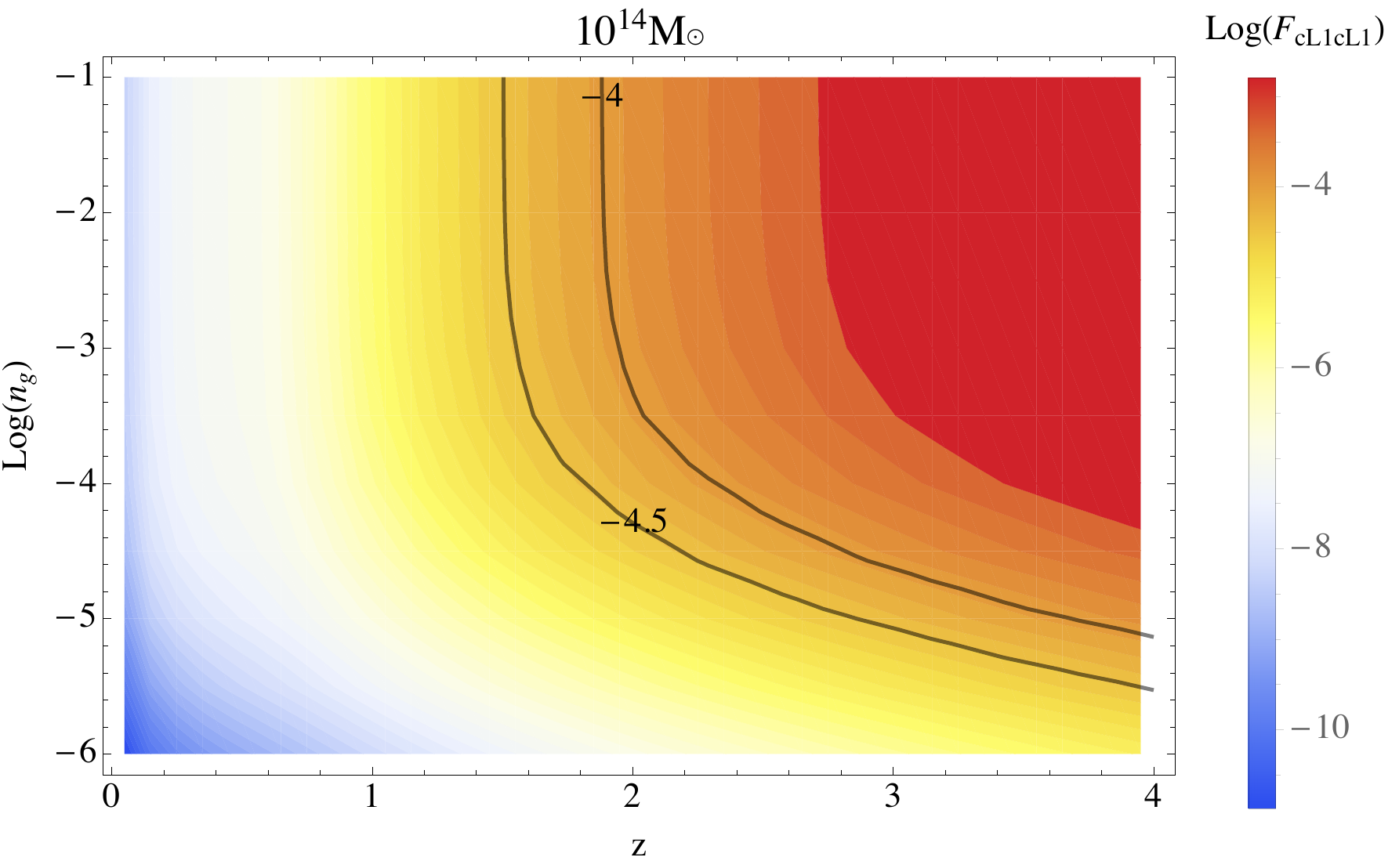}
\caption{Fisher element information $F_{c_{L=1}c_{L=1}}$ as a function of $z$ and the number density $n_g$ in redshift bins of $\Delta z=0.1$, for $f_{sky}=0.75$ and $M=10^{12.5}M_\odot$ (top left panel), $M=10^{13}M_\odot$ (top right panel), $M=10^{13.5}M_\odot$ (bottom left panel) and $M=10^{14}M_\odot$ (bottom right panel).
}
\label{fig:fisher_cL1}
\end{figure*}

In Figure~\ref{fig:fisher_cL1} we plot the Fisher element information $F_{c_{L=1}c_{L=1}}$ as a function of $z$ and the number density $n_g$ in redshift bins of $\Delta z=0.1$, for different halo masses.
As we want to understand if and how it will be possible to reach a precision of $\approx 10$ in the measurement of $c_{L=1}$, we need to reach (in the optimistic hypothesis that we will know the bias from a complementary measurement), $F_{c_{L=1}c_{L=1}} > 0.01$.
It can be seen that redshift bins for $z<2$ don't contribute with a significant amount of information, so one could use at most $\sim 20$ bins, and so each $F$ needs to be of the order of $\approx 5\times 10^{-4}$. In the Figure we plot contours of $Log(F_{c_{L=1}c_{L=1}}) = -4.5, -4$, so we indicate what minimum values of $n_g$ are needed in order to have the required precision.

\begin{figure*}[htb!]
\includegraphics[width=0.49\linewidth]{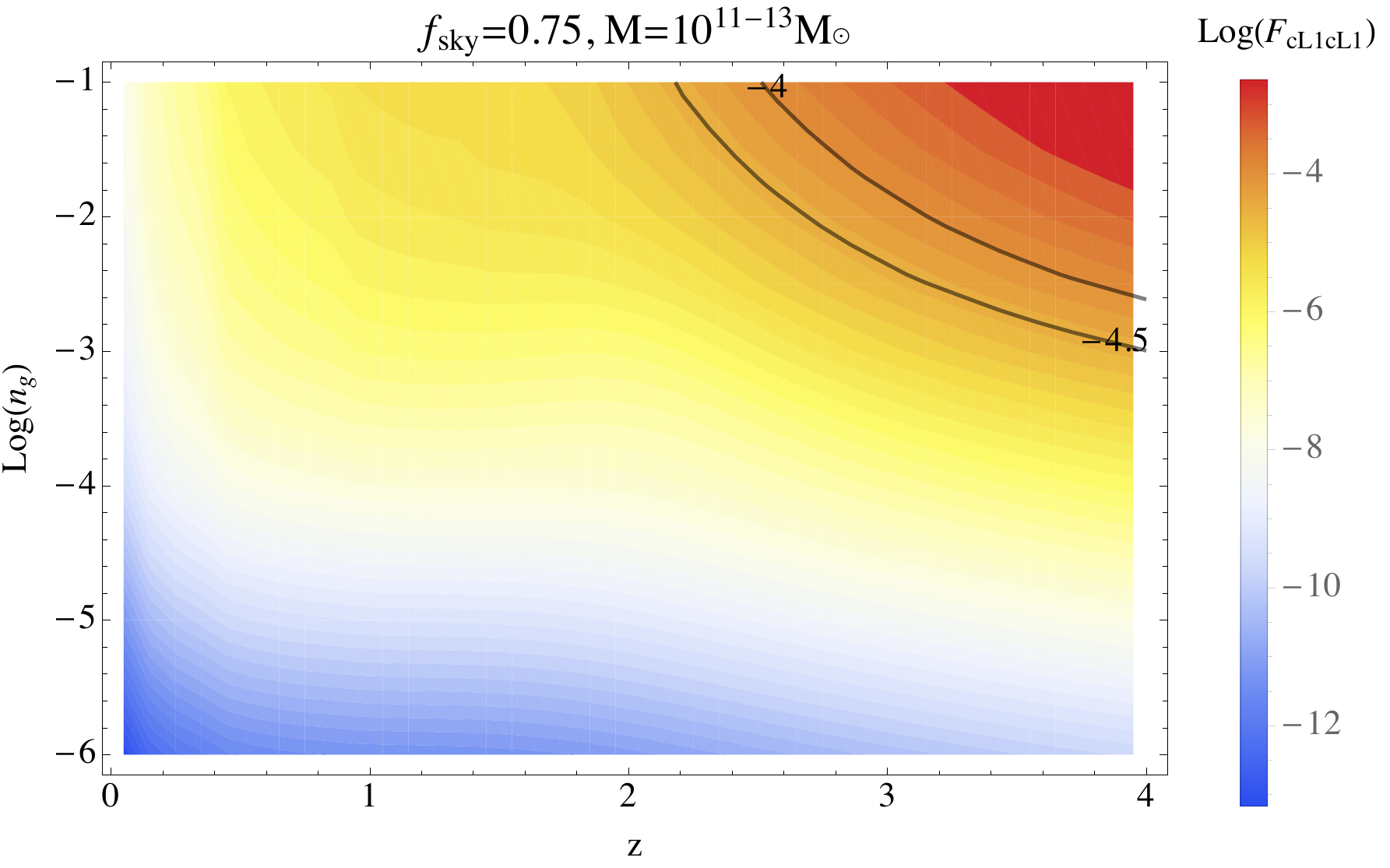}
\includegraphics[width=0.49\linewidth]{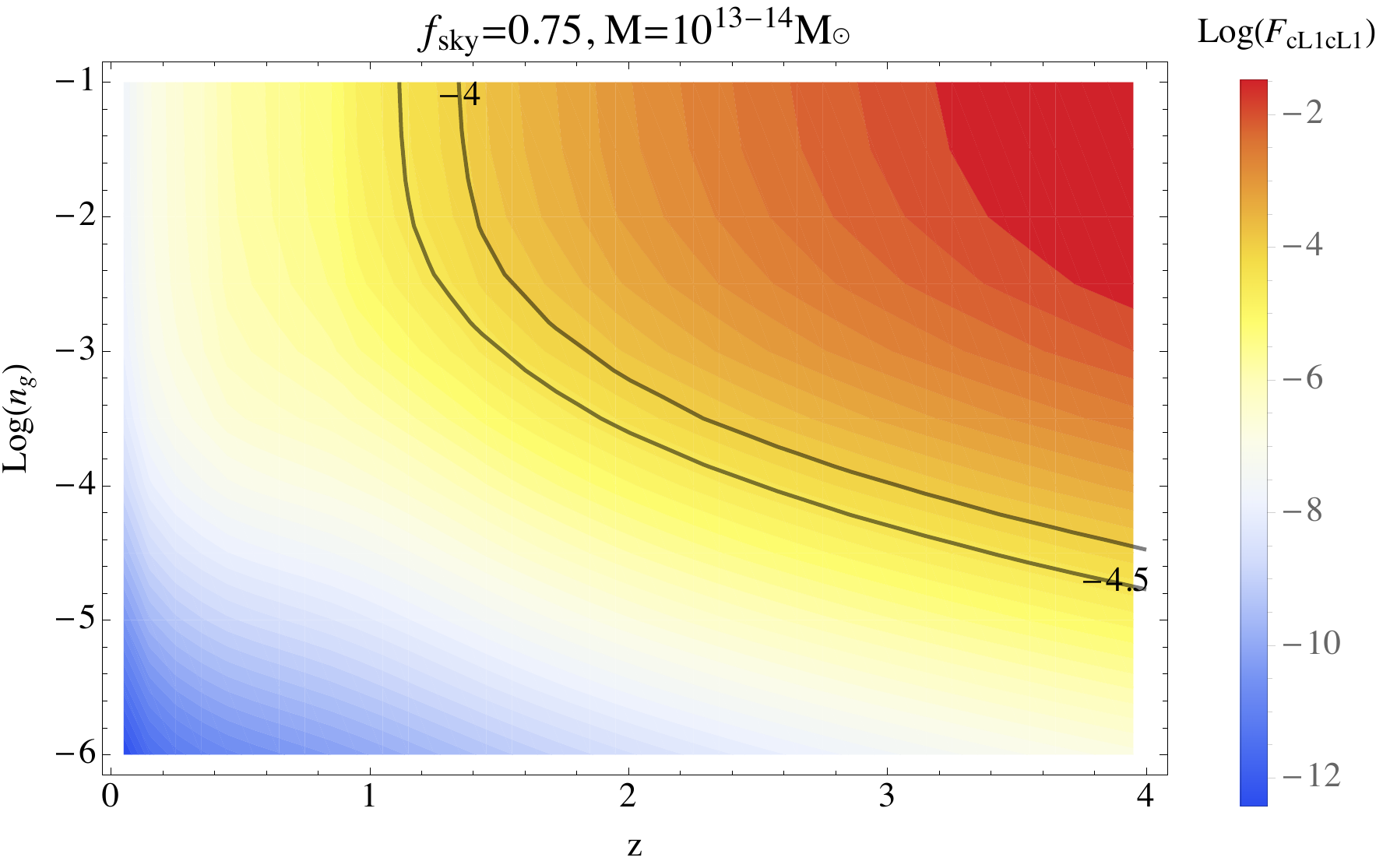}
\includegraphics[width=0.49\linewidth]{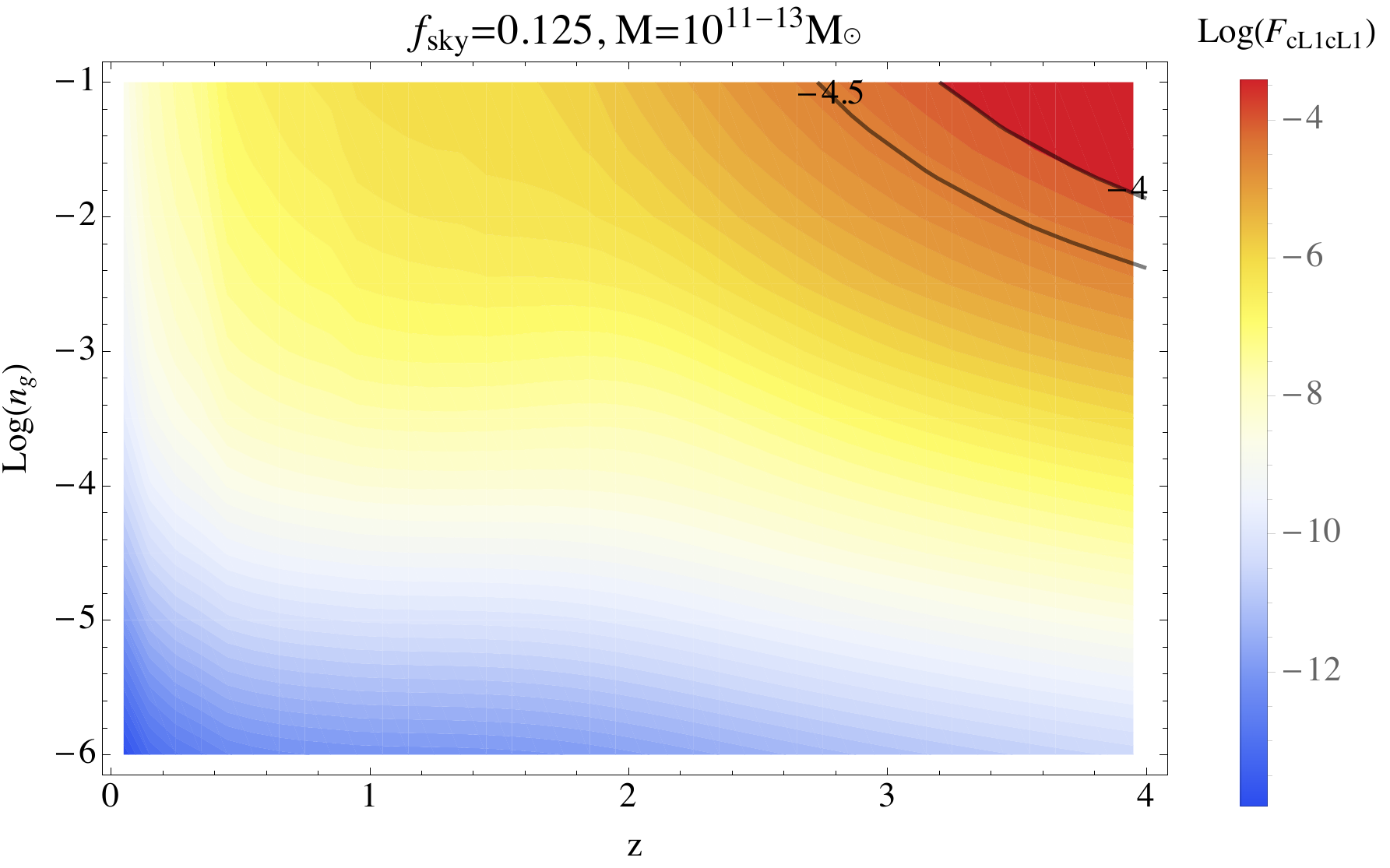}
\includegraphics[width=0.49\linewidth]{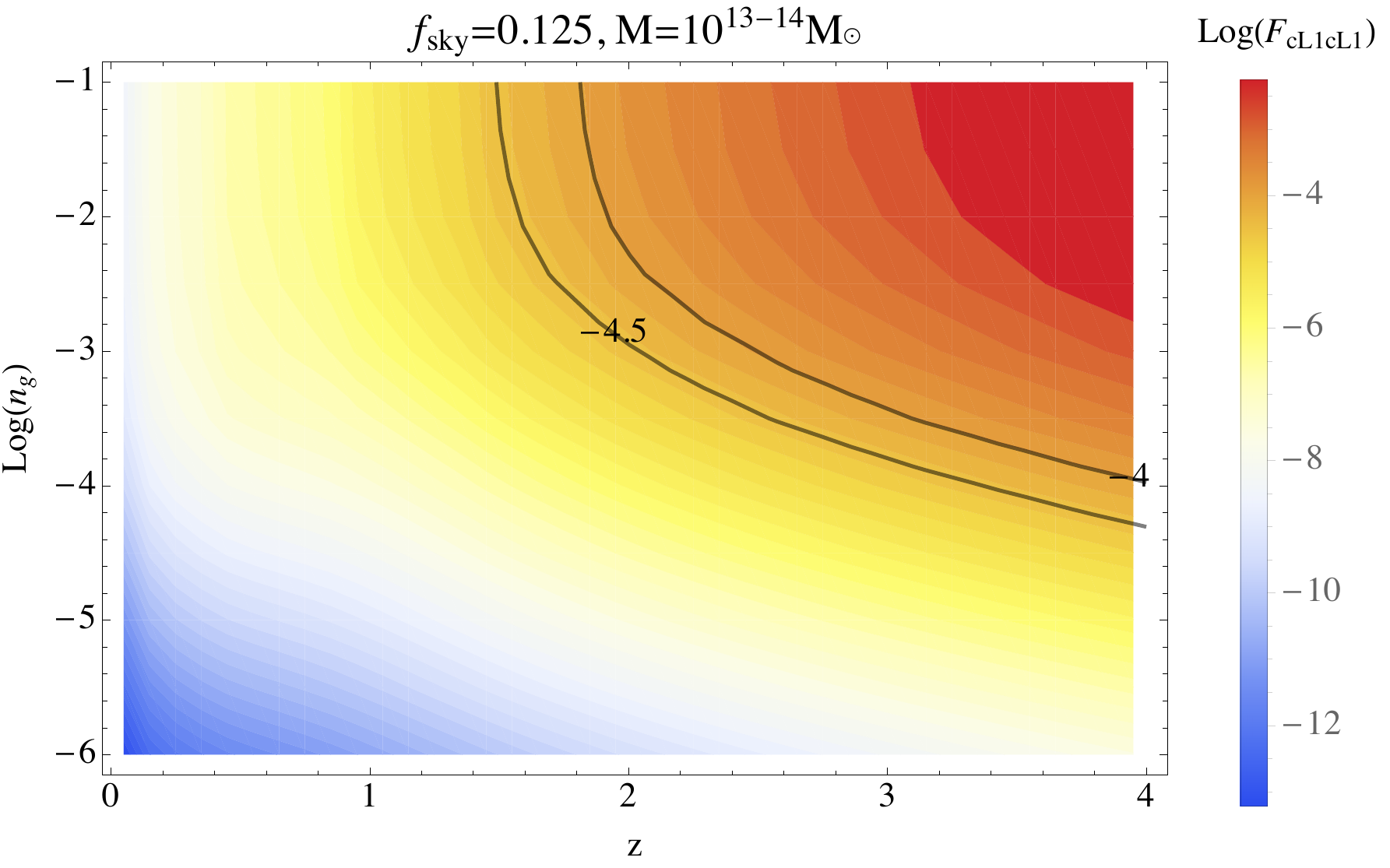}
\caption{Fisher element information $F_{c_{L=1}c_{L=1}}$ for the multi-tracer technique, when assuming $M_1=10^{11}M_\odot$, $M_2=10^{13}M_\odot$, with $n_2=0.01\,n_1$ (left panels), and $M_1=10^{13}M_\odot$, $M_2=10^{14}M_\odot$, with $n_2=0.1\,n_1$ (right panels). Results are for $f_{sky}=0.75$ (upper panels) and $f_{sky}=0.125$ (bottom panels).
Lines represent the minimum number density of sources detected in order to reach $\sigma(c_{L=1})=10$, in bins of $\Delta z =1$.
}
\label{fig:fisher_cL1_MT2}
\end{figure*}

The plots show that it will be difficult to reach the precision we wanted in measuring this parameter with a wide spectroscopic survey, even at high redshift. It is apparent how a very large number of sources at high redshift would be required in order to reach our goal of $\sigma(c_{L=1}) = 10$.
We then investigate a few of examples of the multi-tracer technique and see how much it can be gained with that. We consider the cases of distinguishing between two populations with mass $M_1 < M_2$, and assume that $n_2$ is an order of magnitude smaller for each order of magnitude that $M$ increases.
We show our results in Figure~\ref{fig:fisher_cL1_MT2}, and we can see that the multi-tracer case helps considerably.
It is still required to have a survey with a high number density at high-z, but the requirements are obviously less stringent, and potentially reachable.
In the plots we show, as in the single tracer case, the Fisher element information $F_{c_{L=1}c_{L=1}}$ as a function of $z$ and the number density $n_g$, for two different cases of multi-tracer, for a wide (30,000 deg.$^2$) and a narrower (5,000 deg.$^2$) surveys. We overplot lines showing the minimum number density required for the smaller halo mass population, per redshift bin of $\Delta z=01.$, in order to reach $\sigma(c_{L=1}) \approx 10$.
We can see that for this parameter $f_{sky}$ is not as important as it is for the more studied {\it local} shape $f_{\rm NL}$ (see e.g.~\cite{Raccanelli:2014fNL}), and so a deep survey targeting a large number of highly massive objects would be the easiest way to measure $c_{L=1}$ with high precision.

It is beyond the scope of this paper to investigate details of future surveys and how to reach the required numbers; we leave this as a guide for designing future experiments and a future study.

\section{Conclusions}
\label{sec:conclusions}
In this paper we investigated the possibility of using large-scale radio surveys to measure primordial non-Gaussianity in configurations other than the widely investigated ``local'' shape. In particular, we focused on the possibility of constraining vector field models with the large-scale structure of the Universe. This has not been done so far, and the advent of large galaxy surveys in the next few years makes this timely.

We forecasted the precision on measurements of the primordial non-Gaussanity parameter $f_{\rm NL}$ for the local, equilateral, orthogonal and folded shapes, and the vector models described by the $c_{|L=1,2}$ coefficients, coming from the galaxy power spectrum measured by future galaxy surveys. We computed constraints that will be possible to obtain by having some redshift information added to forthcoming radio continuum surveys such as EMU and the SKA, in different configurations. We found that an optimistic but realistic analysis could allow measurements of the $f_{\rm NL}$ parameter competitive or improving the current limits coming from CMB experiments.
In particular, we found that having the ability to divide the galaxy distribution function into 5 bins, already the coming EMU survey could give constraints better than current Planck limits. If we assume that the distribution can be divided only into 3 bins, the SKA will be needed in order to improve upon CMB measurements.
In a more futuristic perspective, as a proof of principle investigation, we found that LSS analyses could be extremely powerful in measuring non-Gaussianity parameters from any shape, provided that redshift information for a large number of galaxies over a large fraction of the sky and at high-z will be available.
Finally, we studied the specifications required for a generic future galaxy survey in order to reach a precision of $\sigma(c_{L=1})=10$, that would allow test of the primordial Universe models with parity-violating anisotropic sources, e.g., a U(1) vector field coupled to a pseudoscalar and large-scale helical magnetic fields in radiation dominated era.
To achieve an even better precision on measurements of non-Gaussianity parameters, it will be necessary to have a high-redshift wide survey and the use of higher order correlations such as the bispectrum (see e.g.~\cite{ Alvarez:2014, spherex, Munoz:2015}).

\[\]{\bf Acknowledgments:}\\
The authors would like to thank Matt Jarvis and Tamara Davis for useful suggestions and Joyce Byun, Yacine Ali-Ha\"{i}moud, Liang Dai, Marc Kamionkowski, Julian Munoz for interesting discussions.
AR is supported by the John Templeton foundation. MS is supported in part by a Grant-in-Aid for JSPS Research under Grant Nos.~25-573 and 27-10917, and in part by World Premier International Research Center Initiative (WPI Initiative), MEXT, Japan. This work is supported in part by the ASI/INAF Agreement I/072/09/0 for the Planck LFI Activity of Phase E2.
DB acknowledges financial support from the Deutsche Forschungsgemeinschaft through the Transregio 33, ÔThe Dark UniverseÕ.
DP was supported by an Australian Research Council Future Fellowship [grant number FT130101086].


\bibliography{Anisotropic_Inflation_Sub}

\end{document}